\documentclass[
reprint,
superscriptaddress,
altaffilletter,
nofootinbib,
amsmath,
amssymb,
aps,
prd]{revtex4-1}

\usepackage{times}
\usepackage{mathtools}
\usepackage{graphicx}
\usepackage{dcolumn}
\newcolumntype{d}[1]{D{.}{.}{#1}}
\usepackage{bm}
\usepackage{verbatim} 
\usepackage{cprotect} 
\usepackage{url}
\usepackage{subfigure} 
\usepackage{tikz}
\usepackage{afterpage}
\usepackage{nicefrac}
\usepackage{tabularx}
\newcolumntype{L}{>{\raggedright\arraybackslash}X}
\usepackage{multirow}
\usepackage[english]{babel}
\usepackage{natbib}
\usepackage{hyperref}
\hypersetup{
	colorlinks=true,
	citecolor=NavyBlue,
	linkcolor=NavyBlue,
	urlcolor=NavyBlue,
}
\usepackage{aas_macros}
\usepackage{float}

\graphicspath{figures/}

\usepackage[dvipsnames]{xcolor}
\makeatletter
\newif\ifacinbody \acinbodyfalse
\let\ac@keys\@empty
\newcommand{\ac@decl}[3]{
  \expandafter\gdef\csname ac@s@#1\endcsname{#2}
  \expandafter\gdef\csname ac@l@#1\endcsname{#3}
  \ifx\ac@keys\@empty \gdef\ac@keys{#1}\else \g@addto@macro\ac@keys{,#1}\fi}
\newcommand{\ac@S}[1]{\csname ac@s@#1\endcsname}
\newcommand{\ac@L}[1]{\csname ac@l@#1\endcsname}
\newcommand{\ac@emit}[3]{
  \ifcsname ac@done@#1\endcsname
    \ifacinbody \def\ac@out{\hyperlink{ac:#1}{#3}}\else \def\ac@out{#3}\fi
  \else
    \expandafter\gdef\csname ac@done@#1\endcsname{}
    \ifacinbody \def\ac@out{\hypertarget{ac:#1}{\textcolor{NavyBlue}{#2}}}\else \def\ac@out{#2}\fi
  \fi
  \ac@out}
\newcommand{\acbeginbody}{\acinbodytrue
  \@for\ac@k:=\ac@keys\do{\global\expandafter\let\csname ac@done@\ac@k\endcsname\@undefined}}
\DeclareRobustCommand{\gls}[1]{\ac@emit{#1}{\ac@L{#1}\ (\ac@S{#1})}{\ac@S{#1}}}
\DeclareRobustCommand{\glspl}[1]{\ac@emit{#1}{\ac@L{#1}s\ (\ac@S{#1}s)}{\ac@S{#1}s}}
\ac@decl{GW}{GW}{gravitational wave}
\ac@decl{eGW}{eGW}{eccentric gravitational-wave signal}
\ac@decl{eCBC}{eCBC}{eccentric compact binary coalescence}
\ac@decl{CBC}{CBC}{compact binary coalescence}
\ac@decl{BBH}{BBH}{binary black hole}
\ac@decl{SNR}{SNR}{signal-to-noise ratio}
\ac@decl{FAR}{FAR}{false-alarm rate}
\ac@decl{IMR}{IMR}{inspiral--merger--ringdown}
\ac@decl{EOB}{EOB}{effective-one-body}
\ac@decl{PN}{PN}{post-Newtonian}
\ac@decl{NR}{NR}{numerical relativity}
\ac@decl{RR}{RR}{radiation reaction}
\ac@decl{PSD}{PSD}{power spectral density}
\ac@decl{LVK}{LVK}{LIGO--Virgo--KAGRA}
\ac@decl{O3}{O3}{third observing run}
\makeatother

\newcommand{\gstlal}{GstLAL}
\newcommand{\dali}{\texttt{TEOBResumS-Dalí}}
\newcommand{\lalsim}{\texttt{LALSimulation}}
\newcommand{\lalsuite}{\texttt{LALSuite}}

\begin{document}

\title{\bf {More Eccentric, Not Always Less Detectable: High Mass Holds Its Ground, Low Mass Loses Reach}}

\author{Elwin K.~Y. \surname{Li}}
\email{elwinlikayau@link.cuhk.edu.hk}
\affiliation{Department of Physics, The Chinese University of Hong Kong, Shatin, New Territories, Hong Kong}

\author{Juno C.~L. \surname{Chan}}
\email{chun.lung.chan@nbi.ku.dk}
\affiliation{Center of Gravity, Niels Bohr Institute, Blegdamsvej 17, 2100 Copenhagen, Denmark}

\author{Alvin K.~Y. \surname{Li}}
\affiliation{RESCEU, The University of Tokyo, Tokyo, 113-0033, Japan}
\affiliation{Graduate School of Science, The University of Tokyo, Tokyo 113-0033, Japan}

\author{Otto A. \surname{Hannuksela}}
\affiliation{Department of Physics, The Chinese University of Hong Kong, Shatin, New Territories, Hong Kong}

\date{\today}

\begin{abstract}
Gravitational waves from eccentric compact binary coalescences trace dynamical binary formation, but matched-filtering searches responsible for most detections rely on quasicircular templates. Quantifying any sensitivity loss is crucial for assessing population completeness and the need for dedicated eccentric template banks. We present the first systematic \gstlal{} search study spanning eccentricities up to $e_{\rm 15Hz}=0.99$ and broad ranges of chirp mass, mass ratio, inclination, and aligned spin. Using \dali{} waveforms, we inject eccentric signals and measure recovered signal-to-noise ratio, signal-consistency statistics, false-alarm rate, and sensitive volume and distance. For high-mass binaries ($\mathcal{M}_c\gtrsim30\,M_\odot$), sensitive distance is nearly independent of eccentricity, demonstrating that greater eccentricity does not necessarily reduce search sensitivity. In contrast, for low-mass binaries, sensitive distance falls by up to $\sim50\%$ near $e_{\rm 15Hz}\approx0.75$. Although the absolute loss in reach is modest given their low quasicircular sensitivity, this relative reduction exceeds the increase in sensitive distance between the second and third observing runs of the LIGO--Virgo--KAGRA collaboration and is consistent with previous low-mass studies. We also find that waveform matches at fixed source parameters substantially overpredict search sensitivity loss, meaning that the pipeline's signal-consistency test and maximization over its template bank mitigate much of the predicted sensitivity loss. This highlights the need for end-to-end injection campaigns to establish the detectability of eccentric binaries.
\end{abstract}
\maketitle
\acbeginbody

\section{Introduction} \label{sec:introduction}

The detections of \glspl{GW} by the \gls{LVK} 
detectors~\cite{LIGOScientific:2014pky,VIRGO:2014yos,KAGRA:2020tym,LIGOScientific:2021usb,KAGRA:2021vkt,LIGOScientific:2026sit} 
have enabled key constraints on astrophysical properties of compact 
binaries~\cite{LIGOScientific:2018jsj,LIGOScientific:2020kqk,LIGOScientific:2025cwb}.
Although binaries can be circularized by gravitational 
radiation over long timescales~\cite{Peters:1963ux,Peters:1964zz}, 
systems formed through dynamical interactions or entering the detector band at sufficiently small 
separations can retain appreciable 
eccentricity~\cite{Wen:2002km,OLeary:2008myb,Kovetz:2016kpi,Fragione:2017blf,Banerjee:2017mgr,Zevin:2021rtf,Ficarra:2026cej}.
In these scenarios, compact binaries are assembled and hardened through repeated gravitational encounters 
in dense environments, giving rise to subchannels including single-single 
interactions~\cite{OLeary:2008myb,Samsing:2019dtb,Gondan:2020svr}, 
binary-single interactions~\cite{Samsing:2013kua,Samsing:2017rat,Samsing:2017xmd,Rodriguez:2018jqu,Samsing:2020tda,Ginat:2021a}, 
binary-binary interactions~\cite{Zevin:2018kzq}, 
hierarchical interactions~\cite{Wen:2002km,Antonini:2012ad,Antonini:2016gqe,Naoz:2016a,Hoang:2017fvh,Su:2025mdd,Wu:2026hth}, 
gas-assisted mergers~\cite{Tagawa:2019osr,Rozner:2022ydm,Rowan:2025xxb}, 
close encounters in discs~\cite{Samsing:2020tda,Li:2022qoz,Rom:2023kqm,Fabj:2024a}, 
triple systems~\cite{Silsbee:2016djf,Antonini:2017ash,Fragione:2020nib,Martinez:2020lzt,Vigna-Gomez:2020fvw}, 
and von Zeipel–Kozai–Lidov oscillations~\cite{Wen:2002km,Kimpson:2016dgk}.
These mechanisms can efficiently generate binaries with residual eccentricity within the \gls{LVK} 
ground-based detector sensitivity band, in contrast to isolated binary evolution 
pathways~\cite{Dominik:2012kk,Barrett:2017fcw,Giacobbo:2018etu}, 
where prolonged \gls{GW} emission efficiently circularizes the orbit before merger. 
Moderate eccentricities can also significantly increase the detectability of low-frequency 
neutron star tides~\cite{Takatsy:2026doc,Sun:2026lfk}.
Therefore, measuring eccentricity from \glspl{eCBC} provides a direct means to quantify 
the contribution of dynamical formation channels to the observed compact binary 
population~\cite{Zeeshan:2026pga,Rozner:2026jtj,Salvarese:2026klm}.

Considerable effort has been devoted to identifying residual orbital eccentricity in detected compact-binary 
mergers through parameter-estimation 
studies~\cite{Sun:2015bva,Romero-Shaw:2019itr,Wu:2020zwr,Lenon:2020oza,Ramos-Buades:2020eju,Gayathri:2020coq,Romero-Shaw:2022xko,OShea:2021faf,Gamboa:2024hli,Ramos-Buades:2023yhy,Bonino:2022hkj,Iglesias:2022xfc,Fei:2024ruj,Gupte:2024jfe,McMillin:2025hof,Planas:2025feq,Phukon:2025cky,Morras:2025xfu}. 
Several events have shown support for nonzero eccentricity, including 
GW200105\_162426~\cite{Romero-Shaw:2022xko,McMillin:2025hof,Morras:2025xfu,Tiwari:2025fua}, 
although the strength of the evidence remain inconclusive due to coupling with 
waveform systematics, detector noise, the choice of prior in parameter estimation, 
and competing physical effects such as spin precession. 
Consequently, no individual event has yet provided an eccentric interpretation that 
is robust across analysis assumptions. 
Beyond individual events, eccentricity has also begun to be incorporated into population-level 
analyses, where the inferred eccentric fraction can provide information about the relative 
contributions of different compact-binary formation channels~\cite{Gupte:2026whi}.

For the matched-filtering search efforts, however, eccentricity presents a separate challenge. 
Previous studies have suggested that searches based on quasicircular templates can lose sensitivity to eccentric 
signals~\cite{Ramos-Buades:2020eju,Bhaumik:2024cec,Gadre:2024ndy,Vijaykumar:2026pqb}, motivating complementary approaches including weakly modelled methods such as seedless 
clustering~\cite{Tai:2014bfa,Coughlin:2014xqa,Coughlin:2014swa,Tiwari:2015gal} and the construction of eccentric template banks and dedicated eccentric matched-filtering searches~\cite{Phukon:2024amh,Wang:2025yac,j77m-dyym}.
Quantifying the loss in sensitive volume, however, does not by itself identify how 
eccentricity affects the individual components of a matched-filtering search. 
In practice, matched filtering produces a \gls{SNR} time series as each template is 
correlated with the detector data, from which local maxima above a threshold form candidate triggers. 

\begin{figure}[h]
    \includegraphics[width=\columnwidth]{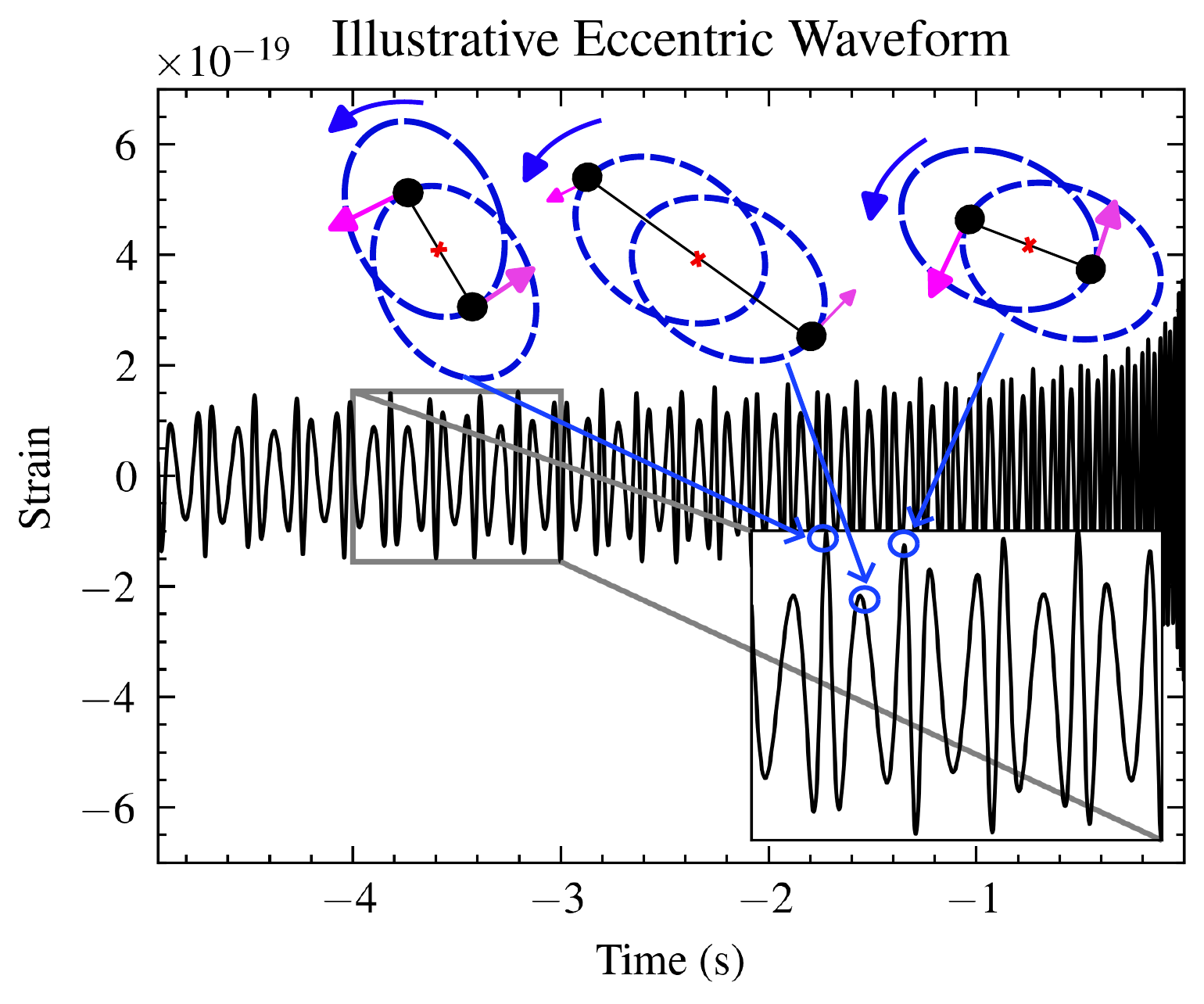}
    \caption{
    Illustrative figure of a strain of a highly eccentric waveform as a function of time. 
    In addition to the usual evolution from inspiral (early time) to merger (late time), 
    the \gls{GW} in the inspiral has peaks and troughs during the close and far encounters 
    in the orbital configuration (see top for illustration). 
    Consequently, highly eccentric waveforms may be amplified such that parts of the waveform 
    inspiral have high signal-to-noise ratio peaks, among other signatures. 
    The waveform morphology is very distinct from quasicircular orbits and may be missed should 
    incorrect templates be used. 
    }
    \label{fig:illustrative_waveform}
\end{figure}

For each triggers, a signal-consistency test value is then evaluated to suppress transient non-Gaussian noise, 
and inconsistencies between the observed matched-filter response and that expected from the template can 
substantially reduce the final ranking of a candidate. 
Waveform mismatch can therefore affect detection efficiency not only through  
alteration in the  recovered \gls{SNR}, but also through its impact on these downstream consistency tests and 
ranking statistics. 
Such an effect has recently been demonstrated for wave-optics distortions induced by gravitational lensing, where 
signals not represented by the template bank can suffer a substantial loss in detection efficiency~\cite{Chan:2024qmb,Chan:2026alt,Wadekar:2026xse}. 
Previous studies have characterized search sensitivity over restricted regions of the eccentric parameter space, 
for example up to eccentricity $e\simeq0.3$ in Ref.~\cite{Ramos-Buades:2020eju} and $e\simeq0.6$ in the search-sensitivity 
analysis of Ref.~\cite{Gadre:2024ndy} using the \texttt{PyCBC}~\cite{Usman2016ThePyCBC} search pipeline. 
We caution that eccentricity is not defined identically across waveform models and reference frequencies, so comparisons 
between our eccentricity at starting frequency 15Hz $e_{\rm 15Hz}$ \cite{Shaikh:2023ypz,Shaikh:2025tae} and the eccentricities quoted in these works are not exact.
A systematic understanding of the search response over a broader eccentricity range is particularly important 
because extending template banks to include eccentricity increases both the computational cost and the number of effective trials. 
The latter can raise the search background and therefore affect the significance assigned to candidate events. 
Establishing where the existing search begins to fail, and which components of the ranking statistic drive this loss, 
is therefore necessary for determining when dedicated eccentric searches are warranted and for guiding the development 
of alternative, computationally efficient techniques to search for these signals.

In this work, we employ the matched-filtering search pipeline 
\texttt{GstLAL}~\cite{Cannon:2012zt,Cannon:2015gha,Messick:2016aqy,Cannon:2021a} 
to systematically investigate how orbital eccentricity affects the detectability of GWs from \gls{BBH}. 
We perform an extensive injection campaign using the \dali{} waveform 
model~\cite{Nagar:2024dzj,Nagar:2024oyk,Albanesi:2025txj}, 
covering a broad parameter space and eccentricities up to $e=0.99$, and recover these signals using 
a standard quasicircular template bank without introducing dedicated eccentric templates. 
This constitutes the first search-sensitivity study employing \dali{} over such a broad eccentricity 
range, extending substantially into the highly eccentric regime, and the first investigation using 
\texttt{GstLAL} to quantify the impact of eccentricity on the search response.
By propagating the injections through the full \texttt{GstLAL} search pipeline, 
we investigate not only the overall loss in detection efficiency but, how 
eccentricity affects different statistics of a matched-filtering search.
In particular, we disentangle the effects of reduced recovered \gls{SNR} 
and changes in the signal-consistency statistic on the final ranking of eccentric \gls{GW} candidates.
Examining these two effects separately, rather than only the overall detection efficiency, 
lets us identify where and why standard quasicircular searches lose sensitivity as eccentricity 
increases, and hence the regimes in which dedicated eccentric templates, or modifications 
to existing search strategies, are needed.

The paper is structured as follows.
Section~\ref{sec:prelim} presents a preliminary study, using the waveform match and the 
signal-consistency statistic to identify the regions of parameter space where eccentricity 
is most likely to affect the search, and to guide the design of the injection campaigns. 
Section~\ref{sec:method} describes the methodology, including the eccentric waveform generation 
and the setup of the injection campaigns in the \gstlal{}~\cite{Cannon:2015gha,Messick:2016aqy,Sachdev:2019vvd} 
matched-filtering search.
Section~\ref{sec:results} presents the results of the injection campaigns, quantifying how the recovered 
\gls{SNR}, the signal-consistency statistic, the false-alarm rate, and the sensitive volume changes with eccentricity.
Section~\ref{sec:conclusion} summarizes our findings and discusses their implications for 
the need for dedicated eccentric searches.
Further details of the \dali{} waveform model and of the \gstlal{} search pipeline are provided in 
Appendix~\ref{appendix:dali} and Appendix~\ref{appendix:gstlal}, respectively.

\section{Motivating the injection campaigns with match and signal-consistency diagnostics} \label{sec:prelim}

To quantify the similarity between \glspl{eGW} and quasicircular gravitational-wave signals, we compute their
waveform matches in Section~\ref{subsec:prelim_match}. 
These matches provide a proxy for the fractional loss in 
optimal matched-filter \gls{SNR} 
and are used to guide the choice of parameter space for the subsequent injection campaigns. 
They also allow us to identify parameters that have negligible influence on the match between 
eccentric and quasicircular waveforms, which can therefore be omitted from the injection 
campaigns to reduce computational cost.

\subsection{Waveform overlap and signal-to-noise ratio}\label{subsec:wf_overlapping}

For two waveforms $h_1(t)$ and $h_2(t)$, the inner product is defined as~\cite{Cutler1994GravitationalWaves}
\begin{equation}
    \langle h_1, h_2 \rangle = 4\,\mathrm{Re} \int_{f_{\mathrm{low}}}^{f_{\mathrm{high}}} 
    \frac{\tilde{h}_1(f)\tilde{h}_2^{*}(f)} {S_n(f)}\,df ,
\end{equation}
where $\tilde{h}(f)$ is the Fourier transform of the gravitational-wave waveform projected onto the detector strain, 
$S_n(f)$ is the one-sided detector noise \gls{PSD}, and 
$[f_{\mathrm{low}},\ f_{\mathrm{high}}]$ define the relevant frequency range for the detector (typically 10 Hz to 2048 Hz). 
To quantify the similarity between an eccentric waveform and the corresponding quasicircular waveform, we compute 
their noise-weighted overlap. 
We estimate the PSD using the third Observing run (O3) of the Advanced LIGO detectors, LIGO Hanford (H1) and LIGO Livingston (L1), 
with GPS time between $1238175530$ and $1238788394$. 
The same PSD is applied consistently across the whole paper. 
All network \gls{SNR} values reported in this work are computed over this two-detector H1--L1 network, 
so that every result is referenced to a common detector sensitivity.
The normalized overlap is then given by
\begin{equation}
    \mathcal{O}(h_1,h_2) = 
    \frac{\langle h_1,h_2\rangle}
    {\sqrt{\langle h_1,h_1\rangle \langle h_2,h_2\rangle}} .
\end{equation}
In practice, the coalescence time $t_c$ and phase $\phi_c$ between the two waveforms are not known a priori. 
We therefore define the match as the overlap maximized over these extrinsic degrees of freedom,
\begin{equation}
    \mathcal{M}(h_1,h_2) = 
    \max_{t_c,\phi_c} \mathcal{O} \left[ h_1, h_2(t_c,\phi_c) \right].
\end{equation}
In principle, one could additionally maximize over the intrinsic parameters of the waveform to and define the fitting factor, 
which can be used to together with the Laplace's approximation in Bayesian methods to estimate detectability while neglecting 
non-Gaussian noise contributions. 
However, since our purpose is to merely identify the regions of parameter space where eccentricity is likely to affect the search, 
we do not perform this additional maximization.

\begin{figure*}[t]
    \centering
    \includegraphics[width=\textwidth]{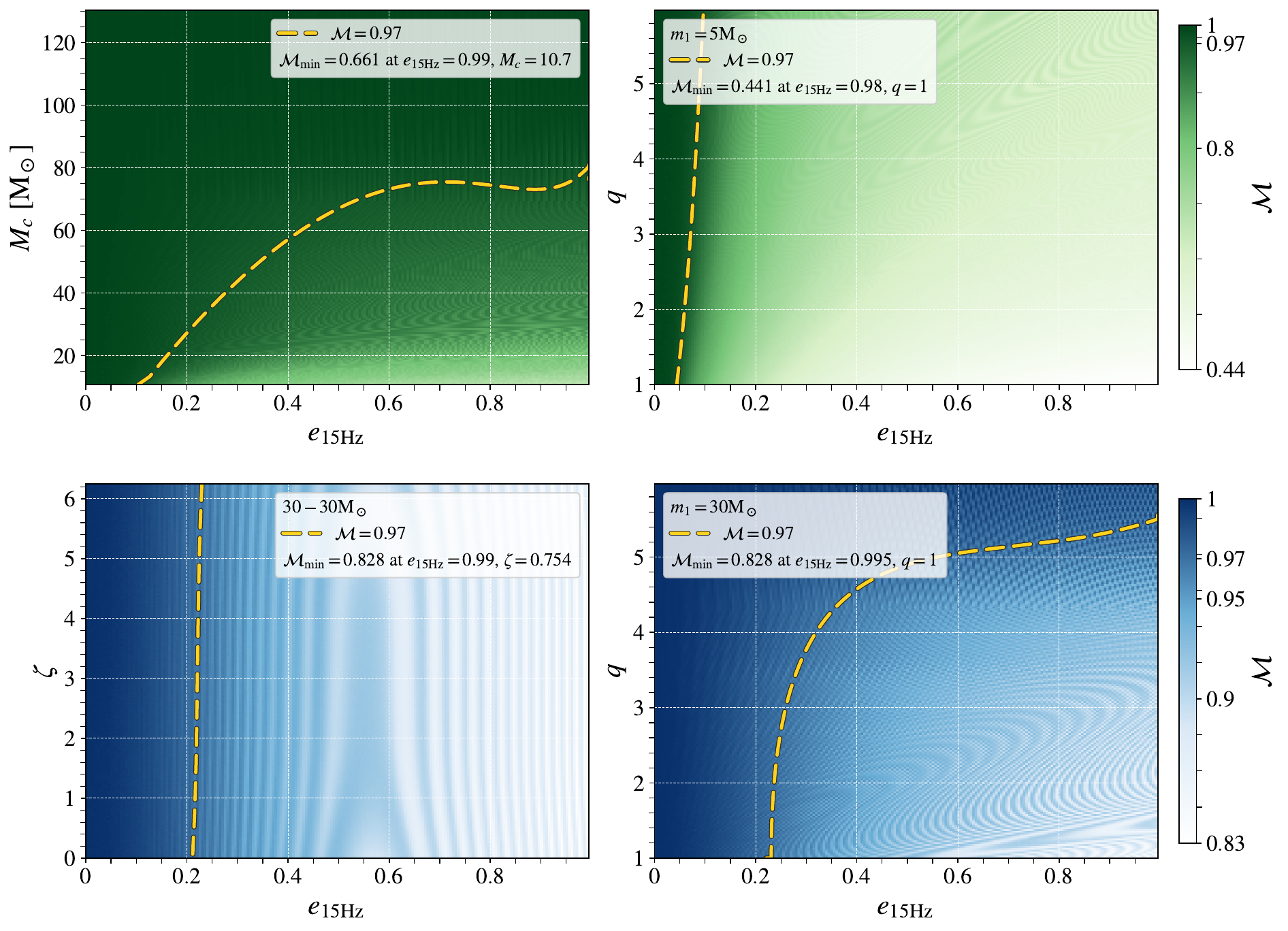}
    \caption{
    Match between eccentric and quasicircular gravitational waveforms, used as a preliminary estimate of 
    the fractional loss in optimal matched-filter \gls{SNR} when eccentric signals are recovered with quasicircular templates. 
    The match is evaluated by varying the eccentricity together with one additional source parameter at a time, 
    while keeping the remaining parameters fixed: 
    chirp mass $\mathcal{M}_c$ (top-left), 
    mass ratio $q$ with $m_1 = 5M_\odot$ (top-right), 
    mass ratio $q$ with $m_1 = 30M_\odot$ (bottom-right), and 
    true anomaly $\zeta$ for a $30-30M_\odot$ binary (bottom-left). 
    Here the mass ratio is defined as $q = m_2/m_1$. 
    Unless otherwise specified by the varied parameter in each panel, the component masses are fixed to $30-30M_\odot$.
    The yellow dashed contours indicate the $\mathcal{M}=0.97$ threshold, corresponding to a $3\%$ mismatch; 
    the contours are smoothed using a polynomial-regression fit. 
    The legend in each panel gives the parameter combination at which the minimum match occurs. 
    These results indicate that eccentricity is one of the dominant parameters controlling the loss in match, 
    with the strongest loss occurring for high eccentricity, lower chirp mass, and nearly equal-mass systems.
    }
    \label{fig:ecc_match}
\end{figure*}

In this work, $h_1$ is taken to be an eccentric waveform generated with \dali{}, while 
$h_2$ is the corresponding quasicircular waveform generated with the same source parameters 
(including the chirp mass, mass ratio, component spins, sky localization, polarization, coalescence time and phase and the source distance), 
except for the orbital eccentricity. 
The match therefore measures the fractional loss in optimal \gls{SNR} caused by neglecting eccentricity in 
the template waveform 
(when not compensating for the bias in the recovered source parameters), 
with a match of $\mathcal{M}=0.97$ corresponding to a $3\%$ loss in optimal matched-filter \gls{SNR}.

\subsection{Optimal match for eccentric signals}\label{subsec:prelim_match}

To choose the region of parameter space that the injection campaigns should target, we compute 
the match between eccentric waveforms and their corresponding quasicircular waveforms over 
a range of eccentricities and use it as a proxy for the fractional loss in 
optimal matched-filter \gls{SNR} incurred when \gls{eGWs} are recovered with quasicircular templates. 
We adopt a match of $\mathcal{M}=0.97$ as the reference threshold for flagging regions where eccentricity 
may cause a non-negligible loss in recovery with quasicircular templates. 
This value follows from the way the template bank is constructed: the bank is laid out so that any target 
signal retains a match of at least $0.97$ with some template, 
a minimal match that bounds the loss in optimal \gls{SNR} to about $3\%$ and the loss in detectable volume 
to roughly $10\%$. 
An eccentric signal whose match with every quasicircular template drops below $0.97$ therefore falls outside the region the bank is 
designed to cover, so $\mathcal{M}=0.97$ marks the approximate level where eccentricity begins to matter for the search. 
By varying the chirp mass, mass ratio and true anomaly separately (Figure~\ref{fig:ecc_match}), 
this proxy lets us identify efficiently where eccentricity most degrades the match and hence which regions of eccentricity, 
chirp mass and mass ratio the dedicated campaigns should concentrate on. 
Here the true anomaly specifies the angular position of the binary along its eccentric orbit relative to periastron, 
which sets the point in the orbit at which the binary enters the detector band.

\subsubsection{Dependence on eccentricity and chirp mass}\label{subsubsec:ecc_vs_cmass}

The dependence of the match on eccentricity and chirp mass is shown in the top-left panel of Figure~\ref{fig:ecc_match}. 
The minimum match (with value $0.661$) occurs in the high-eccentricity and low-chirp-mass region 
(at approximately $e_{\rm 15Hz} = 0.98$ and $\mathcal{M}_c = 4.35M_\odot$). 
This point corresponds to the largest fractional loss in optimal \gls{SNR} among the chirp-mass values considered.

The match varies with chirp mass. 
Indeed, it generally increases with chirp mass, meaning larger chirp-mass systems are less affected by eccentricity, 
mainly because the eccentricity is most visible in the inspiral range when the system has not had significant time 
to circularise. 
For systems with small chirp masses (chirp mass close to $5M_\odot$), the eccentricity is predicted  
to have a significant effect on the search (match decreases noticeably as eccentricity increases). 
However, at higher chirp masses, the match remains high (above 0.97) across the eccentricity range. 
This is because higher-mass systems have fewer in-band cycles in the detector-sensitive frequency band. 
The shorter inspiral leaves less time for eccentricity-induced amplitude and phase modulations to accumulate 
before merger. 
As a result, the eccentric waveform becomes more similar to the corresponding quasicircular waveform, 
leading to less loss in the search sensitivity (a qualitatively higher match).

To save computational resources, it is critical to plan a full injection campaign targeting the regions 
of parameter space where we anticipate the search to be potentially sensitive to eccentricity. 
Therefore, based on these findings, our injection campaigns first focus on binaries moderate- and equal-mass binaries 
with chirp mass $\mathcal{M}_c = 26.1M_\odot$, corresponding to a $30-30M_\odot$ system. 
This choice provides a representative \gls{BBH} configuration while still retaining sensitivity to 
eccentricity-induced waveform differences. 
We deliberately include a second, lower-chirp-mass configuration of around $\mathcal{M}_c = 10.4M_\odot$, 
corresponding to a $12-12M_\odot$ binary, 
where clearly a large sensitivity drop is expected. 
Because lower-mass systems have longer in-band cycles, eccentricity-induced modulations accumulate over more waveform cycles 
and produce the largest match loss among the cases considered. 
This configuration therefore represents the regime in which a quasicircular search is most likely to miss \gls{eGWs}; 
if the search remains sensitive here, it is unlikely to lose them in the more favourable higher-mass regime.
To save computational cost, we do not consider lower-mass systems with chirp mass below $10.4M_\odot$, though for these 
systems the need for dedicated eccentric searches is likely even more pronounced. 

\subsubsection{Dependence on eccentricity and mass ratio}\label{subsubsec:ecc_vs_mr}

We next investigate the effect of eccentricity on the match against mass ratio, 
in particular whether equal-mass or non-equal mass systems are 
the worst case that the campaigns should target. 
The top-right and bottom-right panels of Figure~\ref{fig:ecc_match} show 
the dependence of the match on eccentricity and mass ratio, 
with the first component mass fixed to $m_1 = 5M_\odot$ and $m_1 = 30M_\odot$, 
respectively.

For the lower-mass case with primary mass $m_1 = 5M_\odot$ (top-right panel), 
the match is lowest in the high-eccentricity and equal-mass region 
(with the minimum value $0.441$ occurring at eccentricity of approximately $e_{\rm 15Hz} = 0.98$ and mass ratio $q = 1$). 
Although the match increases mildly with mass ratio, this dependence is relatively weak. 
Once the eccentricity exceeds approximately $e_{\rm 15Hz} \sim 0.1$, the match drops below the $\mathcal{M}=0.97$ threshold 
across nearly the full range of mass ratios considered. 
In the lower-mass regime, therefore, increasing the mass ratio is not expected to 
substantially reduce the loss in optimal \gls{SNR} 
caused by eccentricity. 
This is because the signal duration remains sufficiently long for eccentricity-induced amplitude 
and phase modulations to be imprinted on the waveform.

For the higher-mass case ($m_1 = 30M_\odot$; bottom-right panel of Figure~\ref{fig:ecc_match}), 
the match is again lowest in the high-eccentricity and equal-mass region 
(with the minimum occurring at approximately $e_{\rm 15Hz} = 0.995$ and $q = 1$); 
this because the recurrence of the minimum at a mass ratio of $q=1$ establishes that equal-mass systems 
are the worst case in both mass regimes. 
Unlike the lower-mass case, however, systems with larger mass ratios remain 
more closely matched to their corresponding quasicircular waveforms. 
The practical implication is that, at high mass, a substantial match loss is confined to near-equal-mass systems.
For this reason, we target equal mass binaries in our injection campaigns, which are expected to be the worst 
case for the search.

The mass-ratio dependence at higher mass therefore shows a trend similar to the chirp-mass dependence discussed above. 
In both cases, shorter in-band signal duration reduces the visible impact of eccentricity on the waveform, leading to 
higher matches with quasicircular templates. 
In contrast, equal-mass and lower-mass systems exhibit stronger mismatch at high eccentricity.

\begin{figure*}[t]
    \includegraphics[width=\textwidth]{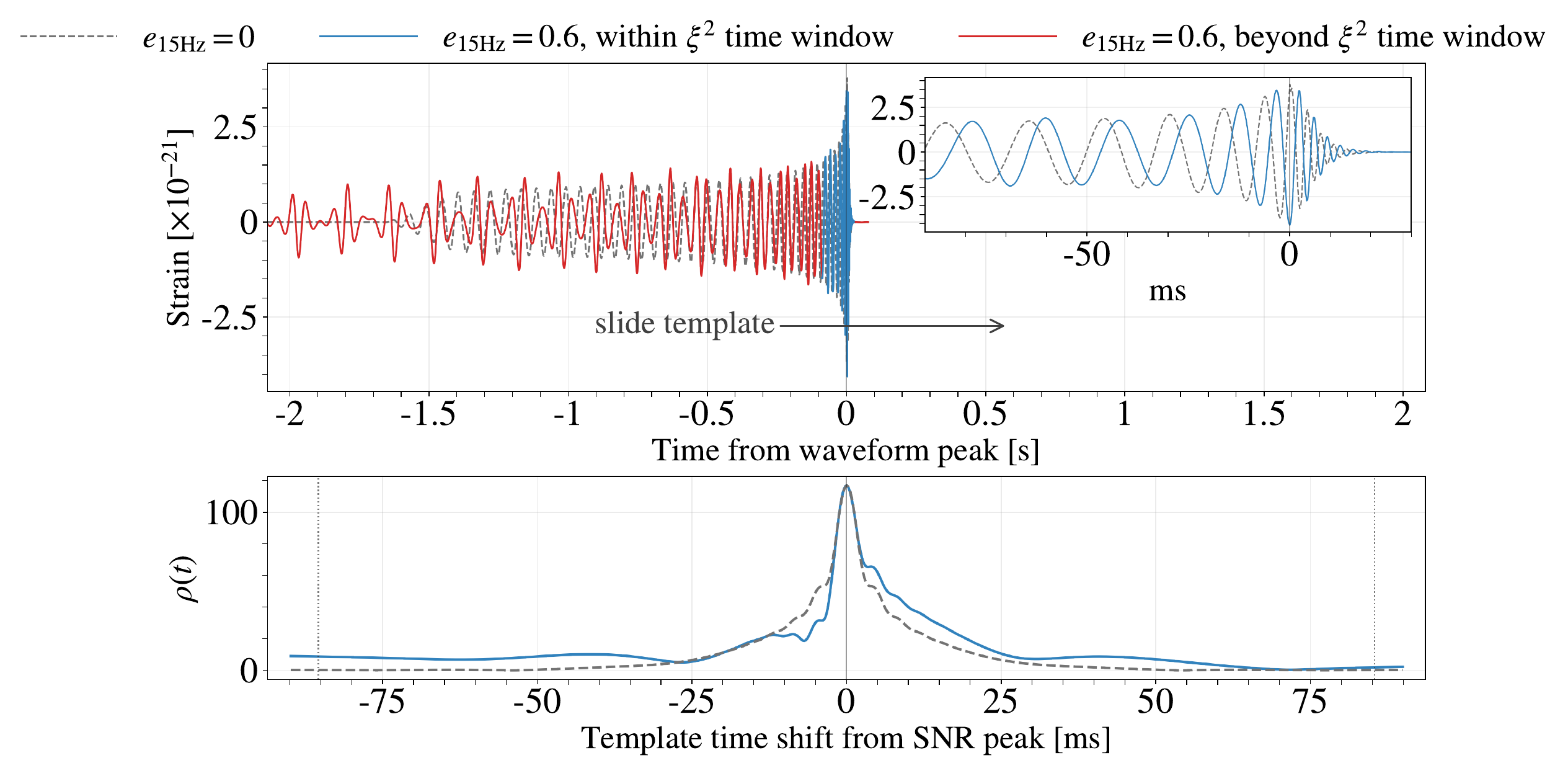}
    \caption{
    Illustration of the autocorrelation-based signal-consistency test used in \gstlal{}. 
    The upper panel shows the time-domain plus polarization for 
    equal-mass $12$--$12\,M_\odot$ waveforms, with the waveform peak aligned at $t=0$. 
    The grey dashed curve is the quasicircular signal, while the coloured curves show an eccentric \dali{} signal with 
    eccentricity $e_{\rm 15Hz}=0.6$. 
    The blue portion indicates the part of the eccentric waveform within the signal-consistency $\xi^2$ time window, 
    while the red portion indicates the part outside this window; 
    the inset zooms into the waveform close to merger. 
    The lower panel shows the corresponding noiseless, \gls{PSD}-weighted matched-filter \gls{SNR} time series obtained by 
    sliding a quasicircular $12$--$12\,M_\odot$ template across each injection. 
    The \gls{SNR} peaks are aligned at zero lag, and the vertical dotted lines mark 
    the $\pm85.4\,{\rm ms}$ window used in the signal-consistency $\xi^2$ calculation. 
    The quasicircular signal gives a normalized signal-consistency statistic $\xi^2/\rho^2 = 2.09\times10^{-34}$, 
    while the eccentric signal gives $\xi^2/\rho^2 = 2.84\times10^{-3}$, 
    showing that eccentricity can produce measurable deviations from the expected template autocorrelation 
    within the merger-centred signal-consistency-test window.
    }
    \label{fig:sample_chisq_plot}
\end{figure*}

Taken together, the preliminary study shows that the match loss is governed primarily by eccentricity and chirp mass, 
is maximized for near-equal-mass systems, 
and depends only weakly on mass ratio away from equal mass. 
This allows the essential parameter space to be covered by a small, targeted set of campaigns rather than a dense grid: 
we fix equal-mass binaries at two representative chirp masses (a $30$--$30\,M_\odot$ and a $12$--$12\,M_\odot$ system) 
and scan the eccentricity up to $e_{\rm 15Hz} = 0.99$. 
Because these configurations sit at the largest match loss found anywhere in the preliminary scan, they constitute 
the worst case for a quasicircular search; 
any conclusion established there, in particular, whether or not the full search actually loses eccentric signals, 
is therefore conservative and carries over to the more favourable regions of parameter space.

\subsubsection{Dependence on eccentricity and true anomaly}\label{subsubsec:ecc_vs_ano}
The last source parameter we check before designing the campaigns is the true anomaly, 
which fixes the orbital phase and hence where in the eccentric orbit the binary 
enters the sensitive band. 
Since scanning it in every campaign would multiply the computational cost, we 
first need to know whether it materially affects the match. 
The bottom-left panel of Figure~\ref{fig:ecc_match} plots the match as a function of eccentricity and true anomaly. 
The match decreases primarily as eccentricity increases. 
This is expected because eccentricity introduces amplitude and phase modulations during the inspiral phase, 
producing waveform features that are not captured by quasicircular templates and therefore reducing the optimal match.

The key result of this slice is that the match varies only weakly with true anomaly compared with its strong dependence on eccentricity. 
Although wave-like structure appears from the phase-dependent alignment between eccentric and quasicircular waveforms, 
the dominant trend remains the decrease in match with increasing eccentricity, 
so within the parameter range studied here entire range of possible true anomaly is a subdominant parameter when it 
comes to search sensitivity. 
This result directly justifies a key design choice: 
rather than scanning the true anomaly, we hold it fixed in the injection campaigns. 
Doing so removes an entire dimension from the campaigns and makes the study computationally tractable, 
without sacrificing coverage of the parameters that actually control the loss.

\subsection{Preliminary signal-consistency test diagnostic} \label{subsec:prelim_chisq}

\gls{SNR} is not the only quantity entering the ranking statistic in 
the \gstlal{} search pipeline. 
The autocorrelation-based signal-consistency statistic, 
$\xi^2$, and the normalized quantity $\xi^2/\rho^2$, 
also play an important role in the search. 
They are used to weed out non-Gaussian noise transients that can mimic \gls{GW} signals. 
Further details of the calculation of the 
signal-consistency statistic $\xi^2$ in \gstlal{} are given in 
Appendix~\ref{appendix:signal_consistency_test}.

Figure~\ref{fig:sample_chisq_plot} illustrates how this signal-consistency test 
is related to the matched-filter \gls{SNR} time series. 
In \gstlal{}, a template is slid across the data to generate an \gls{SNR} 
time series for a trigger. 
This \gls{SNR} time series is then aligned with the expected \gls{SNR} 
time series expected from the autocorrelation of 
the template at the peaks. 
The signal-consistency statistic $\xi^2$ is computed by computing the difference squared between the 
two \gls{SNR} time series within 
a short time window centred on the merger. 
In this example, the window is $\pm 85.4\,{\rm ms}$ around the \gls{SNR} peak. 
Therefore, waveform features outside this merger-centred window 
(red in Figure~\ref{fig:sample_chisq_plot}) 
are not directly included in the signal-consistency $\xi^2$ calculation.

This feature of the \gstlal{} signal-consistency test is important 
when interpreting the injection results. 
Eccentricity-induced modulations can be strongest during the earlier inspiral, 
especially for lower-mass systems with longer in-band durations. 
However, if these modulations occur mainly outside the short signal-consistency $\xi^2$ time window, 
they do not directly contribute to the signal-consistency-test penalty. 
In the full \gstlal{} recovery, triggers are also clustered and ranked 
using the ``likelihood'' that depends on both 
the recovered \gls{SNR} and the signal-consistency information. 
As a result, the recovered trigger is not necessarily the one with 
the largest \gls{SNR} alone, but 
the one that gives the most favourable balance among all the detection statistics. 
This allows the pipeline to recover \gls{eGWs} more effectively than would be expected from 
the preliminary waveform match alone, 
particularly when the strongest eccentric features lie outside 
the merger-centred signal-consistency $\xi^2$ comparison window.

The lower panel of Figure~\ref{fig:sample_chisq_plot} shows the corresponding 
matched-filter \gls{SNR} 
time series for a quasicircular signal and for an eccentric \dali{} signal with 
$e_{\rm 15Hz}=0.6$ generated from $20\,{\rm Hz}$. 
The quasicircular case gives an almost vanishing normalized signal-consistency statistic 
($\xi^2/\rho^2 = 2.09\times10^{-34}$), 
while the eccentric case gives a modest value ($\xi^2/\rho^2 = 2.84\times10^{-3}$). 
This demonstrates that eccentricity can introduce measurable deviations in the 
\gls{SNR} time series relative to 
the template autocorrelation, and hence can affect the signal-consistency statistic $\xi^2$. 
Nevertheless, because the signal-consistency $\xi^2$ calculation is restricted to a short merger-centred window, 
eccentric features outside this window may have a limited effect on the ranking statistic. 

\begin{figure}
    \centering
    \includegraphics[width=\columnwidth]{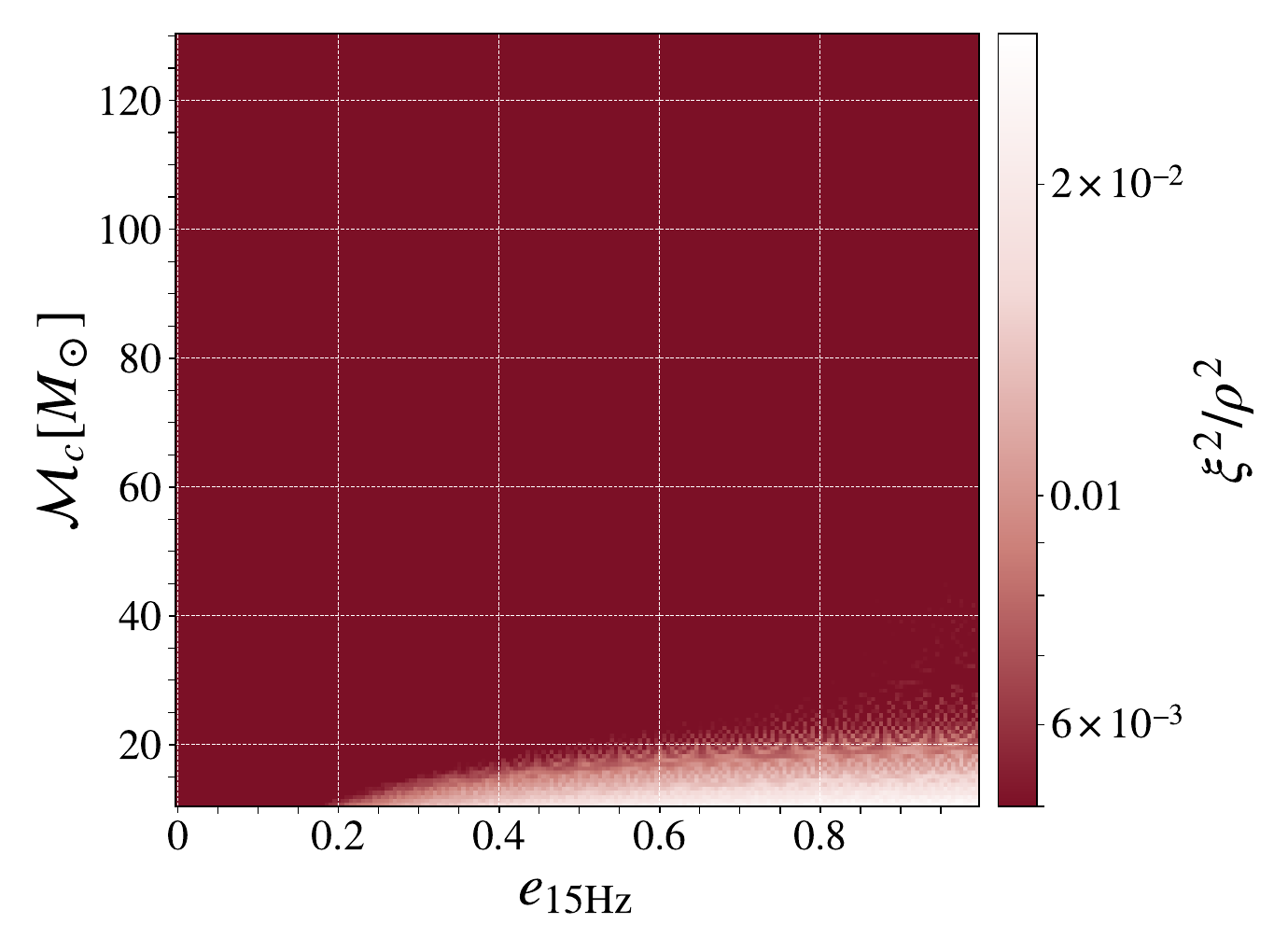}
    \caption{
    Dependence of the normalized signal-consistency-test statistic $\xi^2/\rho^2$ 
    on chirp mass and eccentricity. 
    The signal-consistency $\xi^2$ test is used to weed out 
    non-Gaussian noise transients that can mimic \gls{GW}, 
    but it can also be sensitive to unmodeled residual 
    waveform differences between the true signal (e.g., eccentric waveform strain) 
    and the waveform template (quasi-circular waveform strain). 
    The signal-consistency statistic $\xi^2$ is calculated by comparing 
    the matched-filter \gls{SNR} time series with 
    the expected autocorrelation structure of the template. 
    We normalize the signal-consistency statistic $\xi^2$ by the square of the optimal \gls{SNR}, 
    $\rho^2$, to measure the residual strength relative to the signal strength. 
    The results indicate that eccentricity has little effect on the normalized signal-consistency statistic $\xi^2/\rho^2$ 
    for higher chirp masses, 
    while the statistic increases with eccentricity for lower chirp masses, 
    especially below $\mathcal{M}_c \sim 20\,M_\odot$.
    }
    \label{fig:cmass_ecc_chisq}
\end{figure}

The dependence of the normalized signal-consistency statistic $\xi^2/\rho^2$ on eccentricity and chirp mass is presented in 
Figure~\ref{fig:cmass_ecc_chisq}. 
The largest value occurs at $e_{\rm 15Hz} = 0.99$ and $\mathcal{M}_c = 10.4\,M_\odot$, 
corresponding to a $12$--$12\,M_\odot$ binary. 
For systems with chirp mass below approximately $20\,M_\odot$, 
the value of the normalized signal-consistency statistic $\xi^2/\rho^2$ increases with eccentricity. 
Eccentricity-induced waveform differences therefore become 
more visible to the signal-consistency test in the lower-mass regime.

In contrast, for systems with chirp mass above approximately $20\,M_\odot$, 
eccentricity has little visible effect on the 
normalized signal-consistency statistic $\xi^2/\rho^2$. 
It follows that, at higher masses, eccentric signals can remain sufficiently similar to 
quasicircular templates for the signal-consistency test to produce only small residuals. 
This behaviour is consistent with the preliminary match study: 
higher-mass systems have shorter inspiral durations in the detector-sensitive band, 
leaving less time for eccentricity-induced amplitude and phase modulations to accumulate.

These results provide a preliminary indication that eccentricity is less likely to 
affect search sensitivity for higher-mass systems, while lower-mass eccentric binaries 
may be more vulnerable to reduced sensitivity in quasicircular matched-filtering searches.

Based on the preliminary results (Figures~\ref{fig:ecc_match} and \ref{fig:cmass_ecc_chisq}) 
we select $e_{\rm 15Hz} = 0.4$ and $e_{\rm 15Hz} = 0.99$ as 
two representative eccentricities for the campaigns with $\mathcal{M}_c = 26.1\,M_\odot$. 
Many earlier eccentric waveform models, 
such as TaylorF2Ecc or SEOBNRv4EHM, are valid only for orbital eccentricities 
below approximately $\sim 0.2-0.3$ 
Using the state-of-the-art \dali{} model, which extends to 
very high eccentricities, 
we examine how the number of detected signals changes 
when the eccentricity extends beyond this previously accessible range.

We also carry out injection campaigns at $e_{\rm 15Hz} = 0.99$, 
where highly eccentric signals yield the lowest match with quasicircular waveforms 
and the largest values of the normalized signal-consistency statistic $\xi^2/\rho^2$ 
(Figures~\ref{fig:ecc_match} and \ref{fig:cmass_ecc_chisq}). 
This allows us to test whether the reduction in the optimal match and 
the increase in the signal-consistency-test statistic translate 
into a corresponding 
loss in search sensitivity.

Since the match is largely insensitive to the true anomaly (bottom-left panel of Figure~\ref{fig:ecc_match}) 
we do not vary the true anomaly in these injection campaigns. 
This reduces the computational cost while allowing us to focus on 
the dominant eccentricity dependence.

\section{Methodology}\label{sec:method}

We use \gstlal{}, a time-domain matched-filtering search pipeline, 
to investigate the detectability of \glspl{GW} from compact binaries on 
eccentric orbits in strain-data searches. 
Our aim is to quantify how effectively matched-filtering searches based on 
quasicircular template banks recover eccentric signals, and 
to assess whether eccentricity should be included as an additional parameter 
in template-bank construction. 
Throughout, the injection campaigns use the same two-detector H1--L1 network and
noise \gls{PSD} introduced in
Section~\ref{subsec:wf_overlapping}.
The synthetic eccentric signals are injected into publicly available \gls{O3} strain
data from the two \gls{LVK} detectors, spanning GPS times $1238175530$ to $1238788394$,
and the same stretch of \gls{O3} data is used to estimate the noise \gls{PSD}.
This is the same data epoch analyzed in Ref.~\cite{Chan:2024qmb}, whose injection-campaign
setup we follow, so that the recovery and the preliminary diagnostics are referenced to a
common detector sensitivity.

\subsection{Eccentric waveform model} \label{subsec:generate_wf}

We use \dali{}, implemented in the \lalsim{} package of 
\lalsuite{}~\cite{lalsuite, swiglal}, to generate eccentric waveforms for use as injections. 
This enables the production of eccentric \gls{IMR} waveforms across a broad range of 
orbital eccentricities, including highly eccentric systems. 
We consider binaries spanning orbital eccentricities up to $e \approx 0.99$, 
to examine how detectability depends on orbital eccentricity. 
We note that \dali{} has been validated against numerical relativity at the percent level 
in unfaithfulness only for moderate eccentricities; 
the high-eccentricity ($e_{\rm 15Hz}=0.99$) sets lie beyond the range over 
which such validation currently exists, 
and should be regarded as exploratory stress tests of the search.

For the eccentric waveform generation, we set 
the eccentric reference frequency to $f_{\rm ecc}=15\,{\rm Hz}$ and 
choose this reference point to correspond to periastron. 
This initializes the eccentric orbit such that the waveform begins close to 
the earliest stage at which the eccentric $h_{22}$ frequency evolution can enter 
the LIGO analysis band. 
This choice is particularly important for highly eccentric systems, 
since the instantaneous gravitational-wave frequency varies over the orbit and 
reaches its largest value near periastron.

All injections in this work are generated with 
an initial frequency of $f_0 = 15\,{\rm Hz}$, 
with the reference point chosen to correspond to periastron. 
This value is slightly below, but close to, the typical lower frequency used in 
the analysis band. 
The goal is to assess the detectability of 
eccentric gravitational-wave signals whose eccentricity remains 
measurable within the detector band. 
If the waveforms were generated from a much lower initial frequency, 
the binary could circularize significantly before entering the detector-sensitive band, 
making the observed signal effectively closer to a quasicircular waveform. 
Therefore, $f_0 = 15\,{\rm Hz}$ provides a reasonable choice of initial frequency, 
allowing eccentric effects to remain noticeable before merger without requiring 
an unnecessarily long waveform duration. 
See Section~\ref{sec:prelim} for the detailed rationale behing parameter 
choices and ranges. 

\subsection{Injection campaigns} \label{subsec:campaigns}

We follow a similar injection-campaign setup to Ref.~\cite{Chan:2024qmb} 
and perform five injection campaigns with \gstlal{} to study the impact of eccentricity 
on the search sensitivity to gravitational-wave signals. 
Unless otherwise stated, the eccentric injections are generated using 
the waveform convention described in Section~\ref{subsec:generate_wf}. 
Four of these campaigns are designed to test fixed source configurations: 
two are carried out at fixed chirp mass $\mathcal{M}_c = 26.1\,M_\odot$, 
corresponding to a $30$--$30\,M_\odot$ binary, while the other two are carried out at 
fixed chirp mass $\mathcal{M}_c = 10.4\,M_\odot$, 
corresponding to a $12$--$12\,M_\odot$ binary. 
The final campaign explores the sensitivity volume over a broader source-parameter space. 
We use the preliminary match and signal-consistency-test results in 
Sections~\ref{subsec:prelim_match} and \ref{subsec:prelim_chisq} 
as proxies to identify the eccentricity regions targeted in these 
dedicated injection campaigns.

We also investigate how eccentricity, together with other source parameters, 
affects the sensitivity volume. 
Previous studies~\cite{Gadre:2024ndy} suggest that inclination and spin 
can have a significant impact on the sensitivity volume for eccentric signals. 
Motivated by this, our final injection campaign tests the sensitivity volume 
while varying eccentricity, inclination, and aligned spin for 
two chirp-mass configurations, 
$\mathcal{M}_c = 26.1\,M_\odot$ and $\mathcal{M}_c = 10.4\,M_\odot$. 
This campaign is not intended as a direct comparison with Ref.~\cite{Gadre:2024ndy}, 
but rather as a test of whether similar qualitative behaviour can be reproduced 
using a comparable setup with \gstlal{}.

\subsubsection{Injection campaign 1: eccentricity dependence of search recovery for $30$--$30\,M_\odot$ signals} \label{subsubsec:campaign1}

In this campaign, we explore 
the general detectability of \glspl{eCBC} by 
varying only the eccentricity of the injected signals. 
This provides an initial assessment of how search sensitivity changes with eccentricity. 
It also allows us to test whether recovered \gls{eGWs} signals are down-ranked 
by the \gstlal{} ranking statistic (particularly signal-consistency test) and 
assigned a lower significance relative to background noise.

We systematically sample the eccentricity across $e_{\rm 15Hz} \in [0,\ 0.9]$ 
while holding the remaining source parameters fixed at a 
non-spinning $30$--$30\,M_\odot$ configuration placed at a source distance of 
$125\,{\rm Mpc}$ (see Table~\ref{tab:campaign5_sets} for numerical values).

To isolate the effect on the ranking statistic, we consider a system with little noise. 
We choose the luminosity distance such that the 
optimal injection SNR $\rho_{\rm inj} \sim 100$, and 
randomize the eccentricity across the analyzed data. 
By analyzing the matched-filter \gls{SNR}, detection rates, and 
signal-consistency-test output, we determine where the 
current matched-filtering search pipelines 
are most vulnerable to eccentricity-induced detection loss.

\subsubsection{Injection campaign 2: fixed-eccentricity recovery tests for $30$--$30\,M_\odot$ signals} 
\label{subsubsec:campaign2}

While first campaign studies the general trend of search sensitivity across 
a randomized eccentricity distribution, 
this campaign investigates signal recovery at selected fixed eccentricities. 
To isolate the effect of eccentricity, we construct six sets of injections using 
the \dali{} model, 
each with a fixed eccentricity, 
while keeping all other source parameters unchanged. 
These six eccentricities represent increasing levels of deviation from 
the quasicircular waveform morphology, ranging from the quasicircular baseline 
through low and moderate values, 
which bracket the upper eccentricity ranges accessible to 
several earlier waveform models and 
to the search-sensitivity analysis of Ref.~\cite{Gadre:2024ndy}, 
up to the extreme case $e_{\rm 15Hz}=0.99$, 
where the preliminary match study predicts the largest loss in optimal \gls{SNR} and 
which provides the most stringent test for quasicircular template searches. 
The individual eccentricity values, 
and the fixed source parameters, 
are listed in Table~\ref{tab:campaign5_sets}.

For all injections in this campaign, 
all parameters other than eccentricity are fixed to the same values used in 
the first campaign (Section~\ref{subsubsec:campaign1}). 
This allows us to isolate the impact of eccentricity on signal recovery. 
Matched-filter recovery is then analyzed across the six eccentricity sets 
to assess how increasing eccentricity affects search performance.

The results establish a baseline for the performance of current quasicircular search pipelines 
when recovering eccentric binaries and 
provide insight into the potential need for eccentric waveform templates.

\subsubsection{Injection campaign 3: eccentricity dependence of search recovery for $12$--$12\,M_\odot$ signals} \label{subsubsec:campaign3}

The third campaign repeats the design of the first one, but 
with the component masses changed to lower masses 
(from $30$--$30\,M_\odot$ to $12$--$12\,M_\odot$).
The purpose of this campaign is to test whether the general detectability trends observed 
in the higher-mass case remain valid at lower mass, where eccentricity-induced 
waveform modulations are expected to have a stronger impact.

As in the first campaign, the eccentricity is sampled across 
$e_{\rm 15Hz} \in [0,\ 0.9]$ with the remaining parameters held fixed --- 
here at a non-spinning $12$--$12\,M_\odot$ configuration at the same source distance of 
$125\,{\rm Mpc}$; the corresponding numerical values are listed in 
Table~\ref{tab:campaign5_sets}.

As in the first campaign, 
the eccentricity is randomized across the analyzed data, 
while all other intrinsic parameters are kept fixed. 
We assess whether 
lower-mass eccentric signals are more vulnerable to detection loss 
in quasicircular matched-filtering searches.

We use the same luminosity distance as in the first campaign to be consistent. 
However, since the chirp mass is lower in this campaign, the corresponding 
injected signals are expected to have smaller amplitudes and 
hence lower SNRs at the same distance. 
Consequently, noise fluctuations are expected to have a larger impact on the 
recovered \gls{SNR}, \gls{FAR}, and signal-consistency-tests.

\subsubsection{Injection campaign 4: fixed-eccentricity recovery tests for $12$--$12\,M_\odot$ signals} \label{subsubsec:campaign4}

The fourth campaign is the lower-mass counterpart of the second one. 
We use the same fixed-eccentricity injection sets, 
but change the component masses to $12$--$12\,M_\odot$. 
This allows us to isolate the effect of eccentricity at lower chirp mass 
and to test whether the stronger mismatch predicted by the preliminary studies 
leads to a measurable degradation in search performance.

Matched-filter recovery is then analyzed across the six eccentricity sets and compared with the corresponding $30$--$30\,M_\odot$ results from the second campaign. 
This comparison allows us to determine whether eccentricity has a 
stronger effect on search sensitivity in the lower-mass regime, 
where the longer in-band inspiral duration gives eccentricity-induced amplitude 
and phase modulations more time to accumulate.

\subsubsection{Injection campaign 5: sensitivity volume with varying eccentricity, chirp mass, inclination, and spin} \label{subsubsec:campaign5}

The previous four injection campaigns are designed to isolate 
the effect of eccentricity at fixed source configurations. 
In contrast, fifth campaign investigates how eccentricity affects 
the sensitivity volume when additional source parameters are allowed to vary. 
This campaign is motivated by previous studies~\cite{Gadre:2024ndy}, 
which suggest that inclination and spin can have a significant impact on 
the sensitivity volume for eccentric signals.

In this campaign, we vary eccentricity, chirp mass, inclination, and 
aligned spin using five configurations, 
sampling the eccentricity over $e_{\rm 15Hz} \in [0,\ 0.9]$ in each 
while a different combination of the remaining parameters is either held fixed or varied, 
so that the effect of each parameter on the sensitive volume can be isolated. 
The source parameters of these five configurations, 
together with those of all four preceding campaigns, 
are summarized in Table~\ref{tab:campaign5_sets}.

The purpose of this campaign is to test whether the qualitative dependence of 
sensitivity volume on eccentricity, chirp mass, inclination, and aligned spin 
can be reproduced using \gstlal{} with \dali{} injections. 
Unlike the previous campaigns, which focus on diagnostic quantities 
such as recovered \gls{SNR}, \gls{FAR}, and 
the normalized signal-consistency statistic $\xi^2/\rho^2$ at fixed distances, 
this campaign directly examines changes in the sensitive volume across 
a broader source-parameter space. 
The sensitive volume and distance are estimated by 
Monte-Carlo integration over the injection set (Appendix~\ref{appendix:sensitive_volume}).

\begin{table}[t]
\centering
\begin{tabular}{cc|cccc}
\hline
\hline
Injection Campaign & Set & $e_{\rm 15Hz}$ & $\mathcal{M}_c\,[M_\odot]$ & $\chi_{\rm eff}$ & $\iota$ \\
\hline
1 & i & $[0,\ 0.9]$ & $26.1$ & $0$ & $0$\\
2 & i & $0$ & $26.1$ & $0$ & $0$\\
2 & ii & $0.2$ & $26.1$ & $0$ & $0$\\
2 & iii & $0.4$ & $26.1$ & $0$ & $0$\\
2 & iv & $0.6$ & $26.1$ & $0$ & $0$\\
2 & v & $0.8$ & $26.1$ & $0$ & $0$\\
2 & vi & $0.99$ & $26.1$ & $0$ & $0$\\
3 & i & $[0,\ 0.9]$ & $10.4$ & $0$ & $0$\\
4 & i & $0$ & $10.4$ & $0$ & $0$\\
4 & ii & $0.2$ & $10.4$ & $0$ & $0$\\
4 & iii & $0.4$ & $10.4$ & $0$ & $0$\\
4 & iv & $0.6$ & $10.4$ & $0$ & $0$\\
4 & v & $0.8$ & $10.4$ & $0$ & $0$\\
4 & vi & $0.99$ & $10.4$ & $0$ & $0$\\
5 & i & $[0,\ 0.9]$ & $10.4$ & $[0,\ 0.9]$ & $0$ \\
5 & ii & $[0,\ 0.9]$ & $10.4$ & $0$ & $[0,\ \pi]$ \\
5 & iii & $[0,\ 0.9]$ & $26.1$ & $[0,\ 0.9]$ & $0$ \\
5 & iv & $[0,\ 0.9]$ & $26.1$ & $0$ & $[0,\ \pi]$ \\
5 & v & $[0,\ 0.9]$ & $[10.44,\ 40.44]$ & $0$ & $0$ \\
\hline
\hline
\end{tabular}
\caption{\label{tab:campaign5_sets} 
Summary of the source-parameter configurations of all five injection campaigns. 
Each row corresponds to one injection set, labelled by its campaign number and set index; 
the columns list the eccentricity $e_{\rm 15Hz}$, chirp mass $\mathcal{M}_c$, 
aligned effective spin $\chi_{\rm eff}$, and 
inclination $\iota$. 
A range (e.g. $[0,\ 0.9]$) denotes a parameter varied across the set, 
while a single value denotes a fixed parameter. 
Campaigns~1 and 3 randomize the eccentricity at 
the $30$--$30\,M_\odot$ ($\mathcal{M}_c = 26.1\,M_\odot$) and 
$12$--$12\,M_\odot$ ($\mathcal{M}_c = 10.4\,M_\odot$) configurations respectively; 
Campaigns~2 and 4 fix the eccentricity at discrete values at the same two configurations; 
and Campaign~5 varies the eccentricity together with 
the chirp mass, aligned spin, or inclination to probe the sensitive volume.}
\end{table}

\begin{figure*}[t]
    \centering
    \includegraphics[width=\textwidth]{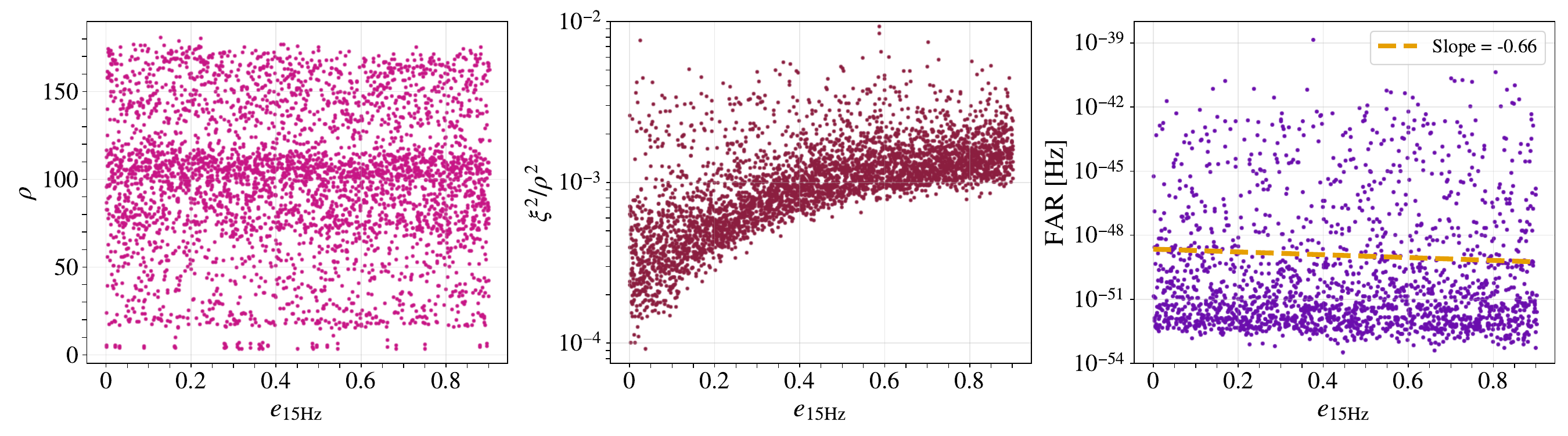}
    \caption{
        Recovery results for injection campaign~1, using $30$--$30\,M_\odot$ 
        eccentric injections with randomized eccentricity. 
        The panels show the recovered network \gls{SNR} $\rho$ (left), 
        normalized signal-consistency-test statistic $\xi^2/\rho^2$ (middle), and 
        false alarm rate (right) as functions of eccentricity. 
        The normalized signal-consistency $\xi^2/\rho^2$ panel is shown on a logarithmic scale. 
        The recovered \gls{FAR} and \gls{SNR} show no clear monotonic degradation with 
        increasing eccentricity; a straight-line fit to 
        $\log_{10}(\mathrm{FAR})$ (dashed line, right panel) 
        has a slope of $-0.66$, negligible against the scatter, 
        confirming the absence of a \gls{FAR} trend with eccentricity. 
        The right panel is restricted to the main low-\gls{FAR} population.
        The normalized signal-consistency statistic $\xi^2/\rho^2$ increases mildly 
        with eccentricity but gradually flattens at higher eccentricities, 
        indicating that eccentricity has only a small effect on the signal-consistency test 
        and does not lead to a significant loss in search sensitivity for this 
        $30$--$30\,M_\odot$ configuration.
    }
    \label{fig:ms3030_vs_ecc}
\end{figure*}

\section{Results} \label{sec:results}

Our results show that, for the $30$--$30\,M_\odot$ mass configuration, 
eccentricity has only a limited effect on the recovery of eccentric signals. 
This is particularly evident in the matched-filter \gls{SNR} and the signal-consistency statistic $\xi^2$. 
Despite the waveform mismatch between eccentric signals and quasicircular templates, 
the overall search sensitivity remains comparable to that obtained 
for quasicircular injections.

For the $12$--$12\,M_\odot$ mass configuration, eccentricity again has only a limited effect 
on signal recovery. 
The matched-filter \gls{SNR} and the signal-consistency statistic $\xi^2$ remain comparable to those 
obtained for quasicircular injections. 
This appears to contradict the expectations from the preliminary match and 
signal-consistency studies (Sections~\ref{subsec:prelim_match} and \ref{subsec:prelim_chisq}). 
We interpret this as a consequence of the clustering 
(that another template can absorb the eccentricity-induced waveform differences) 
and ranking procedure in \gstlal{}, 
which selects triggers by balancing the recovered \gls{SNR} and the signal-consistency statistic $\xi^2$. 
As a consequence of the signal-consistency, 
the recovered trigger is not necessarily the one with the highest \gls{SNR} alone, 
but rather the one that gives the most favourable ranking-statistic value.

In the following subsections, 
we present the results of the five injection campaigns and 
examine the possible reasons for this behaviour. 
We then summarize the implications of these results for 
the general detectability of \glspl{GW} from eccentric binary systems. 
A summary of the injection-campaign results is also provided in Table~\ref{tab:injection_scheme}.

\subsection{Injection campaign 1: eccentricity dependence of search recovery for $30$--$30\,M_\odot$ signals} \label{subsec:results_campaign1}

We compare how the recovered \gls{SNR}, the signal-consistency-test quantity $\xi^2/\rho^2$, and 
the \gls{FAR}, depend on eccentricity. 
Our results show that neither the recovered \gls{SNR} nor the \gls{FAR} exhibits 
a significant dependence on eccentricity (Figure~\ref{fig:ms3030_vs_ecc}).

The left panel of Figure~\ref{fig:ms3030_vs_ecc} shows 
the recovered network \gls{SNR} against eccentricity. 
Since the luminosity distance of the injections is chosen such that 
the estimated injected \gls{SNR} is $\rho_{\rm inj} \approx 100$, 
most recovered injections are concentrated around $\rho \sim 100$. 
Some recoveries scatter above or below this value, 
which is expected due to noise fluctuations in the detector data and 
the clustering procedure in \gstlal{}. 
Because this injected \gls{SNR} is roughly an order of magnitude 
above the detection threshold, these fixed-distance campaigns are not 
designed to probe the detection threshold itself; 
rather, they diagnose the \emph{morphology} of the recovered triggers, 
their distribution in the recovered-SNR--normalized-signal-consistency ($\rho$--$\xi^2/\rho^2$) plane, 
as a function of eccentricity. 
Quantitative sensitivity (sensitive volume and distance) is instead measured in the fifth campagin. 
Again, we do not observe a significant decreasing trend in recovered \gls{SNR} with increasing eccentricity, 
contrary to what might be expected from the preliminary match study (Section~\ref{subsec:prelim_match}). 
This indicates that, at high \gls{SNR}, eccentricity does not strongly shift the 
recovered-\gls{SNR} distribution in the \gstlal{} pipeline.

In the middle panel, 
the signal-consistency-test quantity $\xi^2/\rho^2$ is plotted against eccentricity on a logarithmic scale. 
The values of the normalized signal-consistency statistic $\xi^2/\rho^2$ show 
a mild increase as eccentricity increases, 
indicating that eccentricity introduces a small degradation in the signal consistency 
between eccentric injections and quasicircular templates. 
However, this increase gradually flattens at higher eccentricities, and 
the overall variation remains small compared with the broad recovery behaviour 
shown in the \gls{FAR} and \gls{SNR} panels. 
Eccentricity therefore has a visible 
but weak effect on the signal-consistency test in this $30$--$30\,M_\odot$ campaign, 
without producing a significant loss in search sensitivity.

The right panel presents the \gls{FAR} of the recovered injections as a function of eccentricity. 
Most recovered injections have \gls{FAR} values well 
below the detection threshold $\mathrm{FAR}_{\rm threshold} = 12.1\,{\rm yr}^{-1}$ 
(one false alarm per $30$ days), 
concentrated in a dense low-\gls{FAR} population with a tail extending to higher values from 
injections occurring during noisier detector periods; 
the panel is zoomed to this main population, 
so that a small number of high-\gls{FAR} outliers lying nearer threshold fall outside the shown range. 
To quantify any eccentricity dependence, 
we fit a straight line to $\log_{10}(\mathrm{FAR})$ as a function of eccentricity (dashed line); 
its slope is $-0.66$, negligible against the many-order-of-magnitude scatter in \gls{FAR}. 
We therefore observe no significant trend of \gls{FAR} with eccentricity, 
indicating that, for this high-\gls{SNR} campaign, 
eccentricity does not strongly affect the recovered-\gls{FAR} distribution.

Combining these diagnostic results with the injection-summary results 
shown in Table~\ref{tab:injection_scheme}, 
we conclude that eccentricity has only a limited impact on 
the recovery of $30$--$30\,M_\odot$ eccentric signals 
in this high-\gls{SNR} configuration. 
In particular, the recovered \gls{FAR} and \gls{SNR} show no clear degradation with eccentricity, 
while the increase in the normalized signal-consistency statistic $\xi^2/\rho^2$ remains small. 
Therefore, eccentric signals in this part of the parameter space can still 
be effectively recovered by the \gstlal{} search pipeline using quasicircular templates.

\subsection{Injection campaign 2: fixed-eccentricity recovery tests for $30$--$30\,M_\odot$ signals} \label{subsec:results_campaign2}

As mentioned, 
unlike injection campaign~1, where the eccentricity is randomized continuously across the analyzed data, 
this campaign separates the injections into six fixed-eccentricity sets. 
This allows us to compare the recovery behaviour at specific eccentricities and 
to determine whether highly eccentric signals are systematically down-ranked relative 
to quasicircular signals.

\begin{figure*}[t]
    \centering
    \includegraphics[width=\textwidth]{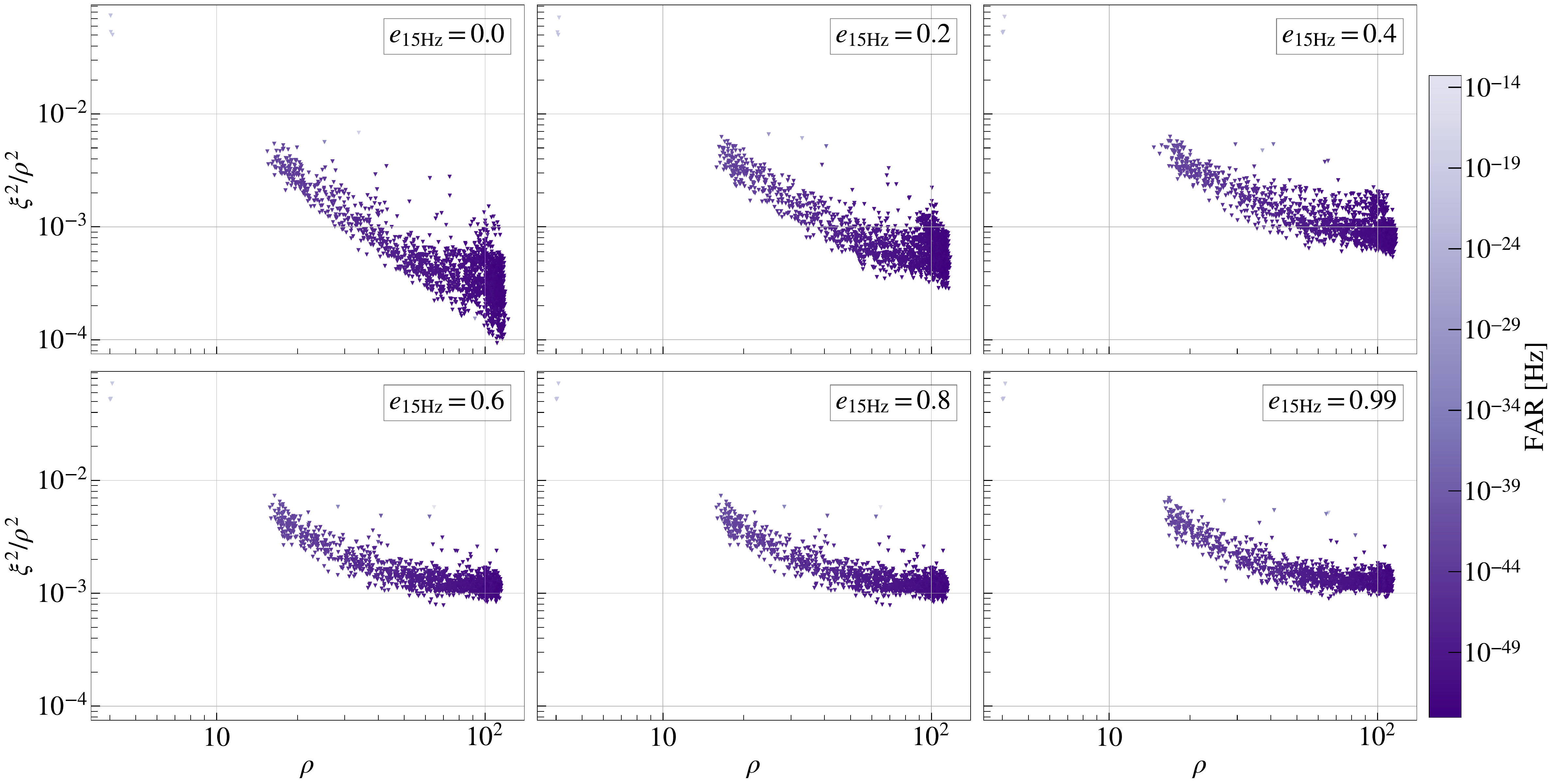}
    \caption{
    Recovered-trigger distribution for the fixed-eccentricity injection sets at $30$--$30\,M_\odot$. 
    Each panel corresponds to one eccentricity value, generated with \dali{}: $e_{\rm 15Hz} = 0$, $0.2$, $0.4$, $0.6$, $0.8$, and $0.99$. 
    The horizontal axis shows the recovered \gls{SNR} $\rho$, while the vertical axis shows the normalized signal-consistency-test statistic $\xi^2/\rho^2$. 
    The colour scale indicates the \gls{FAR} of the recovered trigger.
    Across all eccentricity values, the recovered injections occupy a similar region in the recovered-SNR--normalized-signal-consistency ($\rho$--$\xi^2/\rho^2$) plane, indicating that increasing eccentricity does not produce a clear degradation in recovered \gls{SNR} or signal consistency for this $30$--$30\,M_\odot$ configuration.
    }
    \label{fig:ms3030_output_pass}
\end{figure*}

The recovered-trigger distributions for the fixed-eccentricity injection 
sets at $30$--$30\,M_\odot$ are presented 
in Figure~\ref{fig:ms3030_output_pass}. 
Each panel corresponds to one eccentricity value, 
ranging from the quasicircular case ($e_{\rm 15Hz}=0$) to 
the extreme-eccentricity case ($e_{\rm 15Hz}=0.99$). 
The recovered triggers occupy broadly similar regions in the 
recovered-SNR--normalized-signal-consistency ($\rho$--$\xi^2/\rho^2$) plane across all eccentricities. 
In particular, increasing eccentricity does not lead to a clear systematic decrease in recovered \gls{SNR}, 
nor does it produce a strong increase in the signal-consistency-test statistic. 
The \gls{FAR} values also remain broadly comparable across the six eccentricity sets.

A slight upward shift in the main population of recovered triggers can be observed at higher eccentricities, 
consistent with the weak increasing trend in 
the normalized signal-consistency statistic $\xi^2/\rho^2$ (right panel of Figure~\ref{fig:ms3030_vs_ecc}). 
However, the magnitude of this shift is small. 
It follows that, for the $30$--$30\,M_\odot$ configuration, 
even highly eccentric injections can still be recovered 
with ranking-statistic behaviour similar to that of quasicircular injections.

Therefore, the fixed-eccentricity results are consistent with injection campaign~1: 
eccentricity has only a limited impact on the recovery of high-\gls{SNR} $30$--$30\,M_\odot$ signals 
in the \gstlal{} search pipeline. 
We conclude that, in this part of the parameter space, 
the mismatch between eccentric signals and quasicircular templates 
does not translate into a substantial change in the recovered-trigger 
distribution at this high \gls{SNR}.

\subsection{Injection campaign 3: eccentricity dependence of search recovery for $12$--$12\,M_\odot$ signals} \label{subsec:results_campaign3}

\begin{figure*}[t]
    \centering
    \includegraphics[width=\textwidth]{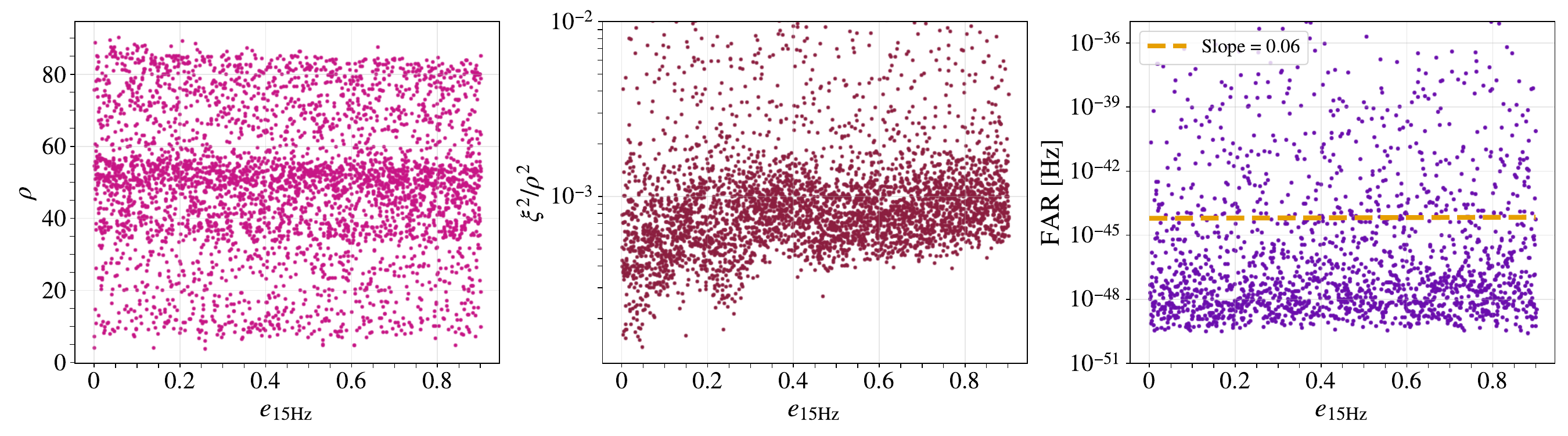}
    \caption{
        Recovery results for injection campaign~3, 
        using $12$--$12\,M_\odot$ eccentric injections with randomized eccentricity. 
        The panels show the recovered network \gls{SNR} $\rho$ (left), 
        normalized signal-consistency-test statistic $\xi^2/\rho^2$ (middle), and 
        false alarm rate (right) as functions of eccentricity. 
        The normalized signal-consistency $\xi^2/\rho^2$ panel is shown on a logarithmic scale. 
        The recovered injections show no clear monotonic degradation 
        in \gls{FAR} or \gls{SNR} with increasing eccentricity; 
        a straight-line fit to $\log_{10}(\mathrm{FAR})$ (dashed line, right panel) 
        has a slope of $+0.06$, consistent with zero within the scatter. 
        The right panel is restricted to the main low-\gls{FAR} population.
        The normalized signal-consistency statistic $\xi^2/\rho^2$ shows only a very weak increasing 
        trend with eccentricity, indicating that eccentricity has a small effect on the 
        signal-consistency test but does not significantly affect the recovery or 
        search sensitivity for this $12$--$12\,M_\odot$ configuration.
    }
    \label{fig:ms1212_vs_ecc}
\end{figure*}

Recall that this campaign tests whether eccentricity 
has a stronger impact on lower-mass systems ($30-30\,M_\odot$ vs $12$--$12\,M_\odot$), 
where the longer in-band inspiral duration allows eccentricity-induced amplitude and 
phase modulations to accumulate over more waveform cycles.
Figure~\ref{fig:ms1212_vs_ecc} presents the dependence of the recovered \gls{SNR}, 
the normalized signal-consistency-test statistic $\xi^2/\rho^2$, 
and the \gls{FAR} on eccentricity. 
The left panel shows the recovered \gls{SNR} as a function of eccentricity. 
Although the recovered \gls{SNR} values show substantial scatter, 
there is no clear systematic decrease as eccentricity increases. 
This implies that, even in the lower-mass configuration, 
eccentricity does not produce a strong degradation in recovered \gls{SNR} in this campaign.

The middle panel plots the variation of the normalized signal-consistency statistic $\xi^2/\rho^2$ 
with eccentricity on a logarithmic scale. 
A very weak increasing trend can be seen as eccentricity increases, 
suggesting that eccentricity has a small effect on the signal-consistency test. 
Notably, this trend is weaker than the corresponding one in 
the middle panel of Figure~\ref{fig:ms3030_vs_ecc} for the high-mass $30$--$30\,M_\odot$ case. 
This difference arises because all four fixed-configuration campaigns 
are injected at the same source distance, 
chosen to keep the comparison between the two mass configurations consistent 
(see Section~\ref{subsubsec:campaign3}). 
At this common distance the lower-chirp-mass $12$--$12\,M_\odot$ injections 
have smaller amplitudes and hence lower recovered SNRs, 
so noise fluctuations have a proportionally larger impact and 
the dependence of the normalized signal-consistency statistic $\xi^2/\rho^2$ on eccentricity is 
less pronounced than in the earlier campaign (Figure~\ref{fig:ms3030_vs_ecc}). 
However, the magnitude of this trend is limited, and 
it does not correspond to a clear degradation in \gls{FAR} or 
recovered \gls{SNR}. 
Therefore, although eccentricity has a modest effect on 
the normalized signal-consistency statistic $\xi^2/\rho^2$, 
the effect is not large enough to significantly impact the recovery or 
search sensitivity for the $12$--$12\,M_\odot$ injections considered in this campaign. 
The right panel shows the \gls{FAR}, 
which likewise exhibits no significant eccentricity dependence: 
a straight-line fit to $\log_{10}(\mathrm{FAR})$ as a function of eccentricity (dashed line) 
has a slope of $+0.06$, consistent with zero within the many-order-of-magnitude scatter. 
As in the $30$--$30\,M_\odot$ case, the panel is zoomed to the main low-\gls{FAR} population.

The right panel shows that the recovered injections are mostly concentrated at very low \gls{FAR} values, 
with a small number of higher-\gls{FAR} outliers. 
These outliers appear across the eccentricity range rather than being concentrated at high eccentricity. 
We do not observe a clear monotonic increase in \gls{FAR} with eccentricity.

\subsection{Injection campaign 4: fixed-eccentricity recovery tests for $12$--$12\,M_\odot$ signals} \label{subsec:results_campaign4}

Injection Campaign~4 is the fixed-eccentricity counterpart of injection campaign~3. 
As in injection campaign~2, the injections are divided into six fixed-eccentricity sets, 
ranging from the quasicircular case $e_{\rm 15Hz}=0$ to the extreme-eccentricity case $e_{\rm 15Hz}=0.99$. 
This allows us to examine whether the recovery behaviour changes systematically 
at specific eccentricity values in the lower-mass configuration.

\begin{figure*}[t]
    \centering
    \includegraphics[width=\textwidth]{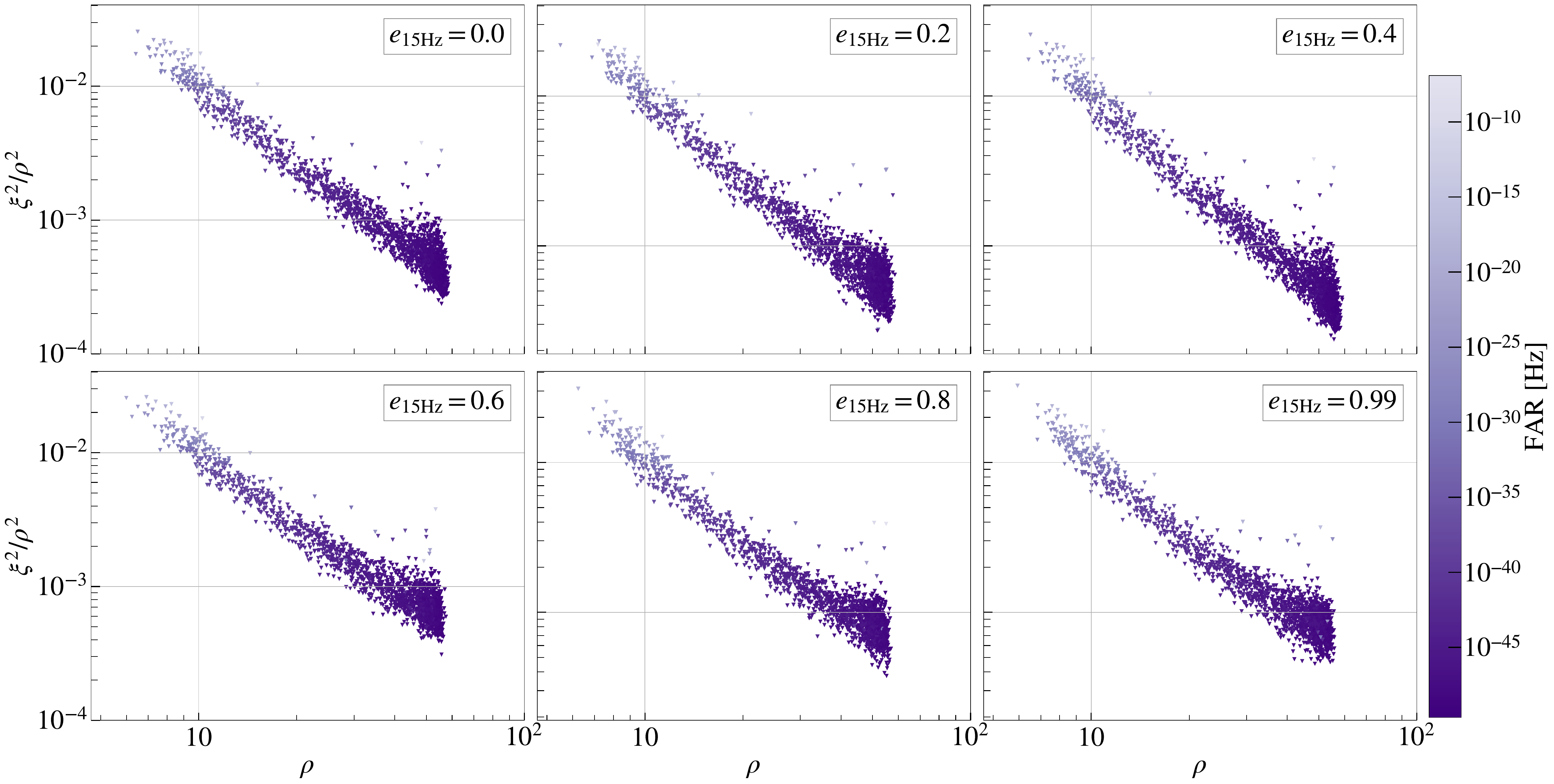}
    \caption{
    Recovered-trigger distributions for injection campaign~4, 
    using fixed-eccentricity  $12$--$12\,M_\odot$ injection sets. 
    Each panel corresponds to one eccentricity value: 
    $e_{\rm 15Hz} = 0$, $0.2$, $0.4$, $0.6$, $0.8$, and $0.99$. 
    The horizontal axis shows the recovered \gls{SNR} $\rho$, 
    while the vertical axis shows the normalized signal-consistency-test statistic $\xi^2/\rho^2$. 
    The colour scale indicates the \gls{FAR} of the recovered triggers.
    The recovered-trigger distributions remain broadly similar across eccentricity values, 
    indicating that increasing eccentricity does not produce a clear systematic degradation 
    in recovered \gls{SNR}, signal consistency, or \gls{FAR} for this lower-mass configuration.
    }
    \label{fig:ms1212_output_pass}
\end{figure*}

The recovered-trigger distributions in 
the recovered-SNR--normalized-signal-consistency ($\rho$--$\xi^2/\rho^2$) plane for 
the six fixed-eccentricity sets are displayed in Figure~\ref{fig:ms1212_output_pass}. 
Across all eccentricities, the recovered triggers occupy broadly similar regions of parameter space. 
The dominant structure is a decrease in the normalized signal-consistency statistic $\xi^2/\rho^2$ with 
increasing recovered \gls{SNR}, 
which is present in all panels and is not unique to the high-eccentricity cases. 
The \gls{FAR} values (colour scale) also show broadly similar behaviour across the eccentricity sets.

There is no clear evidence that increasing eccentricity causes a systematic loss of recovered 
\gls{SNR} or a strong increase in the normalized signal-consistency statistic $\xi^2/\rho^2$. 
Even for $e_{\rm 15Hz}=0.99$, the recovered-trigger distribution remains similar to those obtained at 
lower eccentricities. 
This suggests that, for this lower-mass configuration, 
fixed-eccentricity injections can still be recovered by the \gstlal{} pipeline 
with ranking-statistic behaviour broadly comparable to the quasicircular case.

Combined with the randomized-eccentricity results from injection campaign~3, 
these results indicate that eccentricity has only a limited effect on 
the recovery of lower-mass eccentric signals in the parameter space considered here. 
Although the preliminary match and signal-consistency studies suggest that 
lower-mass systems should be more sensitive to eccentricity-induced waveform differences, 
the full \gstlal{} recovery does not show a strong change in recovered \gls{SNR} 
or signal consistency across the eccentricity range studied at this high \gls{SNR}.

Taken together, the results of the first four injection campaigns show that eccentricity has 
only a limited impact on the recovered-trigger distribution of both the $30$--$30\,M_\odot$ and 
lower-mass eccentric signals considered in this study. 
The recovery statistics for all four campaigns are collected in 
Table~\ref{tab:injection_scheme}, which shows that the fractional loss of recovered injections stays 
at the few-percent level across every eccentricity set, 
with no systematic increase towards high eccentricity. 
We caution, however, that because these injections are placed at $\rho_{\rm inj}\approx100$, which is 
well above the detection threshold, this few-percent loss is set largely by data quality and 
detector duty cycle rather than by eccentricity: 
consistent with this, the loss is slightly \emph{larger} for the $30$--$30\,M_\odot$ sets than for 
the $12$--$12\,M_\odot$ sets, 
the opposite of the trend expected if eccentricity were driving it. 
These fixed-distance campaigns therefore serve as morphology diagnostics, 
and the detectability conclusions of this work rest on the sensitive-volume and 
sensitive-distance measurements of injection campaign~5. 
This appears to contradict the preliminary match study (Section~\ref{subsec:prelim_match}) 
where increasing eccentricity was expected to produce a larger mismatch with quasicircular templates and 
hence a larger loss in optimal \gls{SNR}. 
We interpret this difference as a consequence of the full recovery procedure in \gstlal{}.  
In the preliminary study, the match is computed between fixed eccentric and quasicircular waveforms, 
whereas in the full search the pipeline clusters triggers and selects the recovery that gives 
the most favourable ranking-statistic behaviour, effectively balancing the recovered \gls{SNR} 
against the signal-consistency-test statistic $\xi^2$. 
Therefore, the recovered trigger is not necessarily the one with 
the largest \gls{SNR} alone, but the one that provides the best combination of 
\gls{SNR} and signal consistency.

\begin{figure*}[t]
    \centering
    \includegraphics[width=\textwidth]{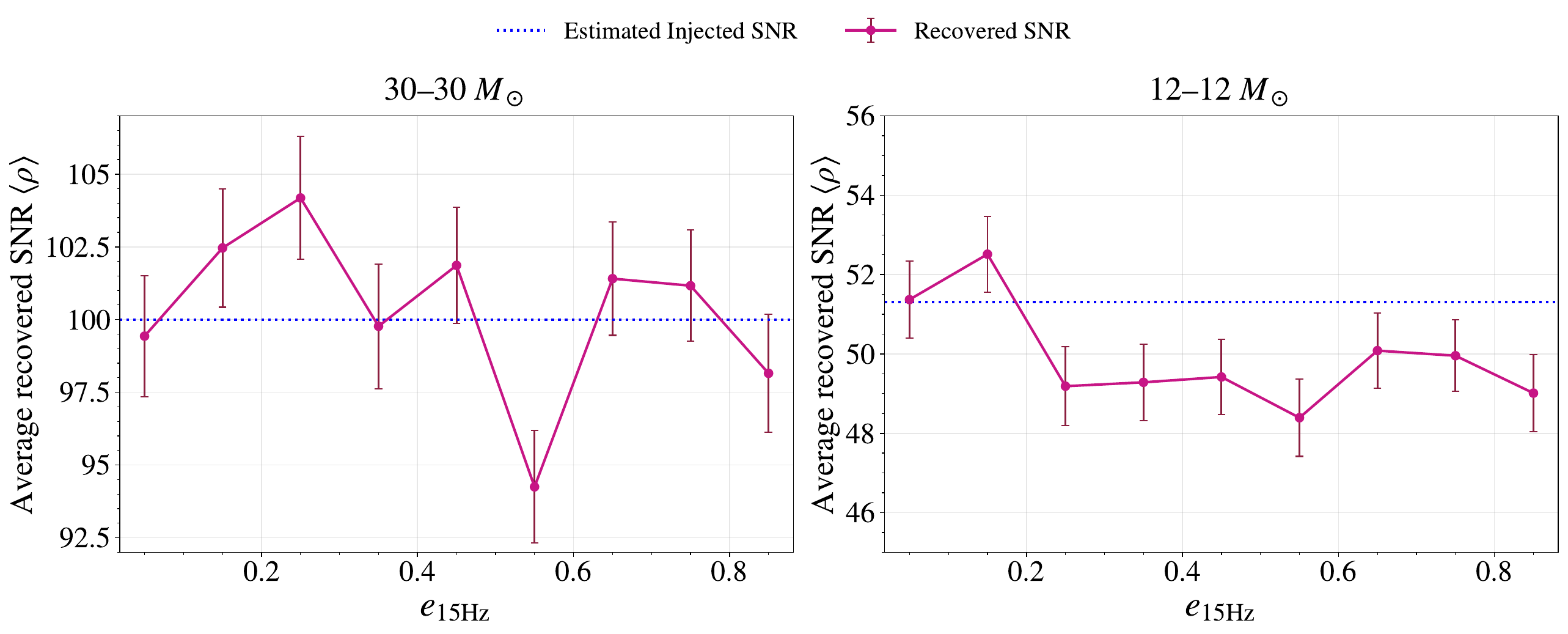}
    \caption{
        Average recovered network \gls{SNR} $\langle\rho\rangle$ as a function of eccentricity for 
        the $30$--$30\,M_\odot$ (left, injection campaign~1) and 
        $12$--$12\,M_\odot$ (right, injection campaign~3) configurations.
        Each point is the mean recovered \gls{SNR} in an eccentricity bin, and 
        the blue dotted line marks the estimated injected \gls{SNR}.
        The error bars quantify the uncertainty in the mean recovered \gls{SNR} 
        after accounting for the antenna response and 
        the modeled detector effects; 
        they do \emph{not} include a correction for waveform mismatch, 
        which can only systematically reduce the recovered \gls{SNR}.
        The horizontal line therefore represents a reference mean, 
        not an upper limit on the recovered \gls{SNR}: 
        bin averages may lie above or below it because the injection samples and detector responses differ, 
        and the error bars quantify the uncertainty in each recovered mean rather than the 
        expected difference from the reference.
        A deficit relative to the reference provides evidence for mismatch, 
        subject to the accuracy of the reference and the uncertainty model.
    }
    \label{fig:avg_snr}
\end{figure*}

\begin{figure*}[t]
    \centering
    \includegraphics[width=\textwidth]{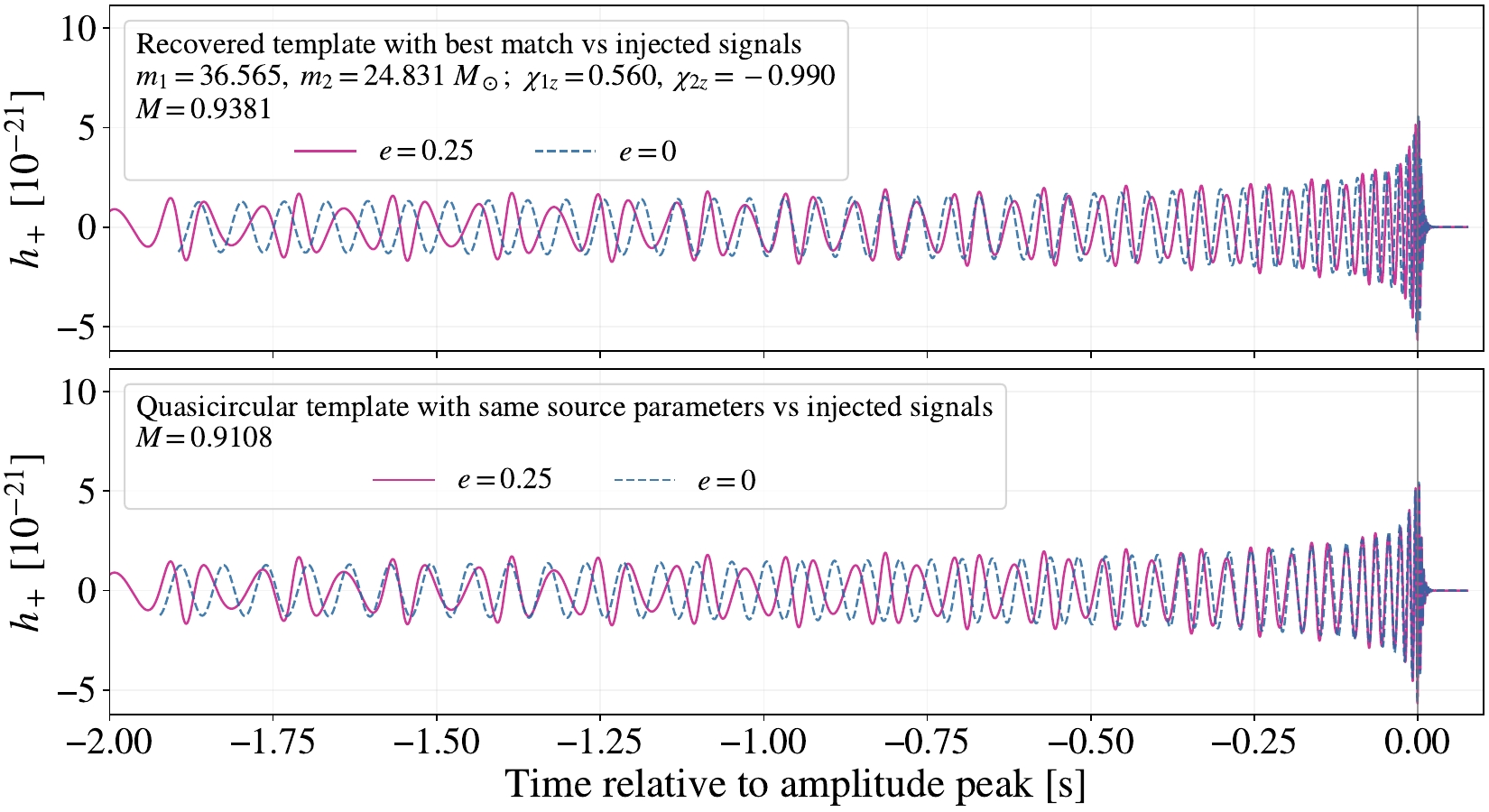}
    \caption{
        Time-domain plus polarization $h_+$ of a representative eccentric injection compared with 
        two quasicircular templates, aligned at the amplitude peak. 
        In both panels the pink solid curve is the eccentric injection ($e=0.25$) and 
        the blue dashed curve is a quasicircular ($e=0$) template.
        \emph{Upper panel}: the best-matching template found by the search, 
        whose intrinsic parameters 
        ($m_1 = 36.565\,M_\odot$, $m_2 = 24.831\,M_\odot$, $\chi_{1z} = 0.560$, $\chi_{2z} = -0.990$) 
        differ from those of the injection; it achieves a match of $\mathcal{M} = 0.9381$.
        \emph{Lower panel}: a quasicircular template with the \emph{same} source parameters as the injection 
        (except for zero eccentricity), which achieves a lower match, $\mathcal{M} = 0.9108$.
        Because the search maximizes over the whole template bank, it recovers the eccentric signal 
        with the higher-overlap best-matching template rather than the same-source quasicircular one, 
        which is why the recovered \gls{SNR} can exceed 
        a same-source reference (Figure~\ref{fig:avg_snr}).
    }
    \label{fig:wf_overlap}
\end{figure*}

To make the comparison between the two mass configurations more quantitative, Figure~\ref{fig:avg_snr} shows the average recovered network \gls{SNR} in each eccentricity bin for both the $30$--$30\,M_\odot$ (injection campaign~1) and $12$--$12\,M_\odot$ (injection campaign~3). The error bars give the uncertainty on each recovered mean after accounting for the antenna response and the modeled detector effects; because they do not correct for waveform mismatch, which can only reduce the recovered \gls{SNR}, a mean lying below the reference provides evidence for mismatch. For the $30$--$30\,M_\odot$ configuration the bin averages fluctuate around the injected reference, scattering both above and below it with no systematic deficit and with error bars broadly consistent with the reference, confirming that eccentricity does not systematically reduce the recovered \gls{SNR} at high mass. For the $12$--$12\,M_\odot$ configuration the recovered mean is close to the reference at low eccentricity but sits slightly below it for $e_{\rm 15Hz} \gtrsim 0.2$; this small but consistent deficit provides mild evidence for eccentricity-induced mismatch in the lower-mass regime. The size of the deficit is nonetheless modest and, as the recovery and \gls{FAR} results show, does not translate into a significant loss of detections.

A notable feature of Figure~\ref{fig:avg_snr} is that several eccentricity bins, 
particularly in the $30$--$30\,M_\odot$ configuration, have a mean recovered \gls{SNR} that lies 
\emph{above} the estimated injected value. 
This does not imply that more \gls{SNR} is recovered than is present, 
but reflects two effects. 
First, the horizontal line is a nominal reference rather than the true optimal \gls{SNR} of each injection: 
the actual optimal \gls{SNR} depends on the detector noise power spectral density at the injection time and 
on the antenna response, both of which vary, so the true injected \gls{SNR} scatters around 
the reference and a fraction of injections genuinely lie above it. 
Second, and more importantly, \gstlal{} maximizes the matched-filter output over the 
entire template bank and over time, retaining the loudest trigger. 
Since the bank is dense, the search can recover the eccentric injection with 
a template whose intrinsic parameters differ from those of the injection 
but which matches the eccentric waveform better than the ``correct'' same-parameter quasicircular template. 
Figure~\ref{fig:wf_overlap} illustrates this explicitly: 
for a representative eccentric ($e = 0.25$) injection, the best-matching template found by the search, 
with shifted component masses and spins, achieves a match of $\mathcal{M} = 0.9381$, 
whereas the quasicircular template with the same source parameters as the injection achieves 
only $\mathcal{M} = 0.9108$. 
By selecting the best-matching template out of the whole bank, 
the search recovers more \gls{SNR} than a same-source reference would imply, 
which is why the recovered means can lie above the nominal injected value even 
though no \gls{SNR} is created.

\begin{table*}[t!]
\begin{tabular}{cc|ccccc|ccc}
\hline
\hline
Injection Campaign & Set & $e_{\rm 15Hz}$ & Source distance [Mpc] & Chirp Mass [$\,M_\odot$] & $\chi_{\rm eff}$ & $\iota$ [rad] & No. injections & Found injections & \% loss [\%]\\
\hline

1 & i & $[0,\ 0.9]$ & $125$ & $26.1$ & $0$ & $0$ & $1973$ & $1898$ & $3.80$\% \\
2 & i & $0$ & $125$ & $26.1$ & $0$ & $0$ & $1973$ & $1901$ & $3.65$\% \\
2 & ii & $0.2$ & $125$ & $26.1$ & $0$ & $0$ & $1973$ & $1900$ & $3.70$\% \\
2 & iii & $0.4$ & $125$ & $26.1$ & $0$ & $0$ & $1973$ & $1899$ & $3.75$\% \\
2 & iv & $0.6$ & $125$ & $26.1$ & $0$ & $0$ & $1973$ & $1900$ & $3.70$\% \\
2 & v & $0.8$ & $125$ & $26.1$ & $0$ & $0$ & $1973$ & $1900$ & $3.70$\% \\
2 & vi & $0.99$ & $125$ & $26.1$ & 0 & 0 & $1973$ & $1897$ & $3.85$\% \\
3 & i & $[0,\ 0.9]$ & $125$ & 10.4 & 0 & 0 & $1949$ & $1915$ & $1.74$\% \\
4 & i & $0$ & $125$ & $10.4$ & $0$ & $0$ & $1949$ & $1913$ & $1.85$\% \\
4 & ii & $0.2$ & $125$ & $10.4$ & $0$ & $0$ & $1949$ & $1916$ & $1.69$\% \\
4 & iii & $0.4$ & $125$ & $10.4$ & $0$ & $0$ & $1949$ & $1912$ & $1.90$\% \\
4 & iv & $0.6$ & $125$ & $10.4$ & $0$ & $0$ & $1949$ & $1914$ & $1.80$\% \\
4 & v & $0.8$ & $125$ & $10.4$ & $0$ & $0$ & $1949$ & $1913$ & $1.85$\% \\
4 & vi & $0.99$ & $125$ & $10.4$ & $0$ & $0$ & $1949$ & $1915$ & $1.74$\% \\

\hline
\hline
\end{tabular}
\caption{\label{tab:injection_scheme} 
Injection-recovery summary for injection campaigns~1--4. 
Each row corresponds to one injection set, labelled by its campaign number and set index, 
and lists the source parameters (eccentricity $e_{\rm 15Hz}$, source distance, 
chirp mass $\mathcal{M}_c$, aligned effective spin $\chi_{\rm eff}$, and inclination $\iota$) 
together with the total number of injections, the number recovered (``found''), and 
the corresponding fractional loss. 
An injection is counted as found when the \gstlal{} false-alarm rate of its recovered trigger falls 
below the threshold $3.85\times10^{-7}\,{\rm Hz}$ 
(see Sec.~\ref{subsec:gstlal_likelihood} for the \gls{FAR} definition). 
Campaigns~1 and 2 use the $30$--$30\,M_\odot$ ($\mathcal{M}_c = 26.1\,M_\odot$) configuration and 
Campaigns~3 and 4 the $12$--$12\,M_\odot$ ($\mathcal{M}_c = 10.4\,M_\odot$) configuration; 
Campaigns~1 and 3 randomize the eccentricity, while Campaigns~2 and 4 fix it at discrete values. 
The recovery loss remains at the few-percent level for every set, 
with no systematic trend with eccentricity. }
\end{table*}

For the lower-mass cases, the discrepancy can also be understood from the waveform morphology and 
from how the signal-consistency statistic is evaluated. 
Although lower-mass systems have longer in-band signal durations and eccentricity is expected to be 
more visible during the early inspiral, the binary can circularize during the inspiral before 
reaching the late-inspiral and merger regime, 
where a large fraction of the matched-filter \gls{SNR} is accumulated. 
Moreover, the signal-consistency statistic $\xi^2$ is computed using a relatively short time window 
concentrated around the merger, typically about $85.4\,{\rm ms}$. 
By this stage, the eccentricity-induced amplitude modulations may already be substantially reduced. 
As a result, the \gls{SNR} time series near merger can remain similar to that produced by 
quasicircular templates, so the signal-consistency statistic $\xi^2$ is not strongly increased. 
Consequently, the search sensitivity does not degrade as much as the 
preliminary match study alone would suggest.

\begin{figure*}[t]
    \centering
    \includegraphics[width=\textwidth]{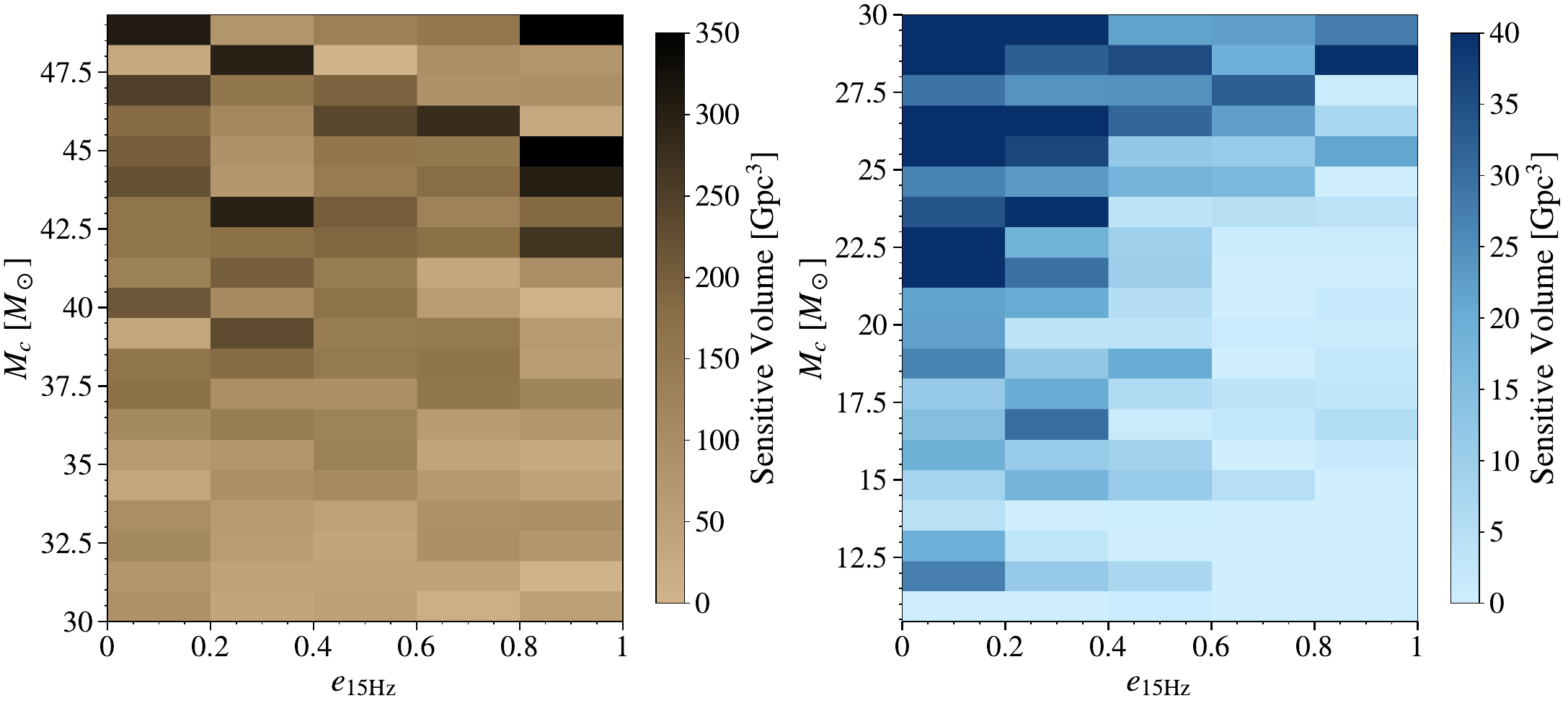}
    \caption{
    Sensitive volume as a function of eccentricity and chirp mass for injection campaign~5, 
    evaluated at a false-alarm-rate threshold of $\mathrm{FAR}_{\rm threshold} = 12.1\,{\rm yr}^{-1}$.
    The horizontal axis shows the orbital eccentricity, while the vertical axis shows 
    the chirp mass $\mathcal{M}_c$.
    The colour scale indicates the sensitive volume in $\mathrm{Gpc}^3$.
    For $\mathcal{M}_c \gtrsim 30\,M_\odot$ the sensitive volume is essentially independent of eccentricity, 
    whereas below $\mathcal{M}_c \approx 30\,M_\odot$ a slight decrease of sensitive volume with increasing 
    eccentricity is visible.
    This mass-dependent trend is consistent with the sensitive-distance results 
    (Figure~\ref{fig:sensitive_distance_far12}) and 
    with the reduced low-mass sensitivity reported in Ref.~\cite{Ramos-Buades:2020eju}.
    }
    \label{fig:sensitivity_volume_cmass}
\end{figure*}

\begin{figure*}[t]
    \centering
    \includegraphics[width=\textwidth]{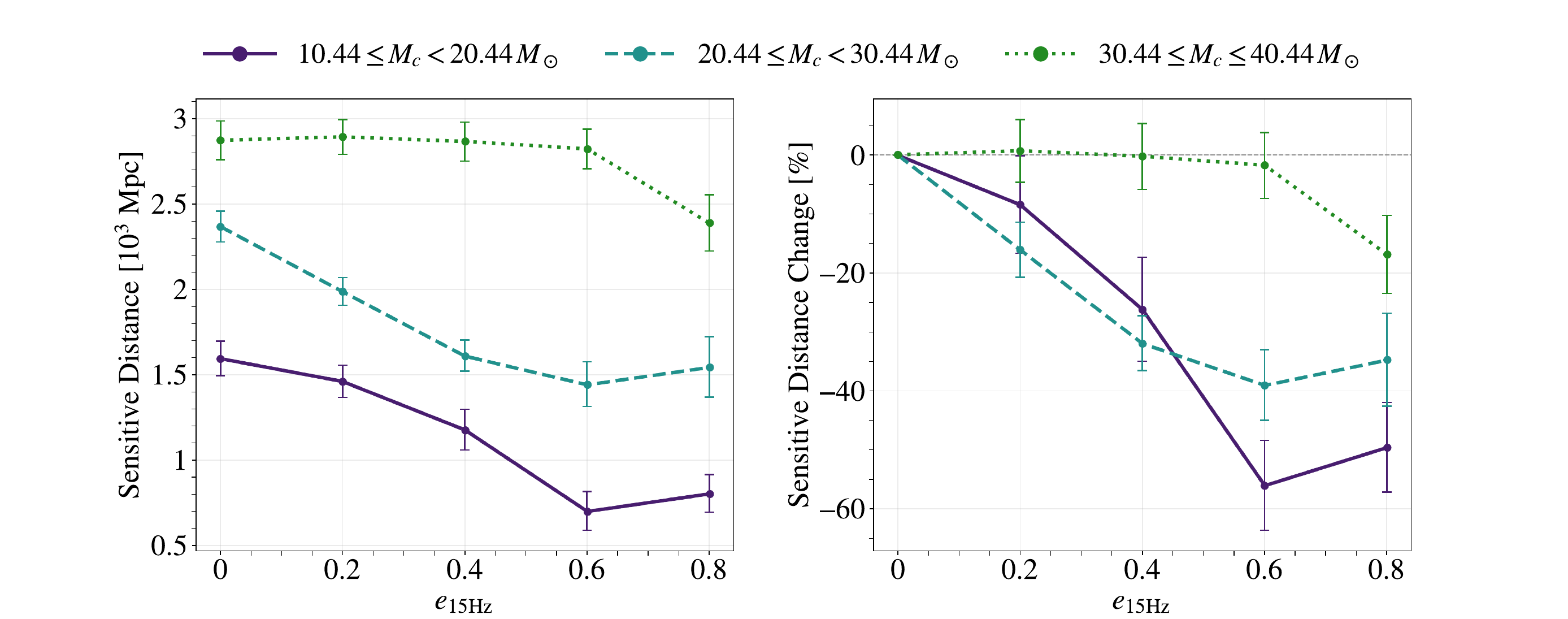}
    \caption{
    Sensitive distance as a function of orbital eccentricity $e_{15\rm Hz}$ for three chirp-mass bins, 
    evaluated at a false-alarm-rate threshold of $\mathrm{FAR}_{\rm threshold} = 12.1\,{\rm yr}^{-1}$ 
    (one false alarm per $30$ days), 
    the standard \gstlal{} detection threshold.
    The left panel shows the absolute sensitive distance in $\mathrm{Mpc}$, and 
    the right panel shows its fractional change relative to the quasicircular 
    ($e_{15\rm Hz} = 0$) case.
    The three curves correspond to the chirp-mass bins 
    $10.44 \le \mathcal{M}_c < 20.44\,M_\odot$ (purple), $20.44 \le \mathcal{M}_c < 30.44\,M_\odot$ (teal), and $30.44 \le \mathcal{M}_c \le 40.44\,M_\odot$ (green).
    The error bars quantify the uncertainty in the estimated sensitive distance after accounting for 
    the antenna response and the modeled detector effects, 
    computed in the same way as for Figure~\ref{fig:avg_snr}.
    For the high-mass bin the sensitive distance is essentially independent of eccentricity, 
    confirming that the effect of eccentricity is negligible at high mass.
    For the two lower-mass bins (both with $\mathcal{M}_c < 30.44\,M_\odot$), 
    the sensitive distance decreases systematically with eccentricity, 
    reaching a reduction of roughly $50\%$ and $40\%$ respectively by $e_{15\rm Hz} \approx 0.75$.
    Because the quasicircular sensitive distance in these lower-mass bins is 
    already comparatively small, however, 
    this large fractional reduction corresponds to only a modest absolute loss in search reach.
    }
    \label{fig:sensitive_distance_far12}
\end{figure*}

\begin{figure*}[t]
    \centering
    \includegraphics[width=\textwidth]{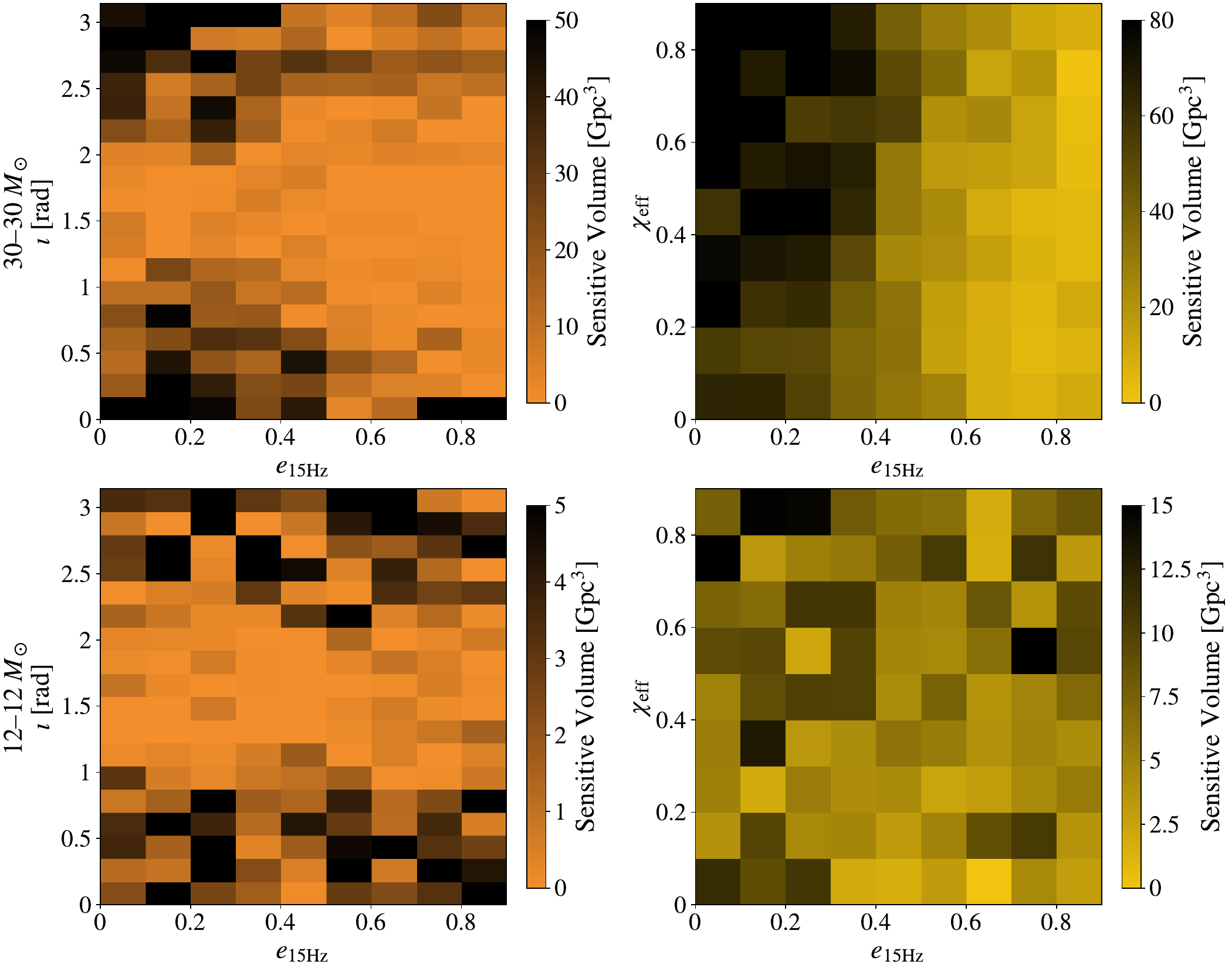}
    \caption{
    Sensitive volume as a function of eccentricity, inclination, and aligned effective spin for 
    injection campaign~5, evaluated at a false-alarm-rate threshold of 
    $\mathrm{FAR}_{\rm threshold} = 12.1\,{\rm yr}^{-1}$.
    The top row shows the $30$--$30\,M_\odot$ configuration, 
    while the bottom row shows the $12$--$12\,M_\odot$ configuration.
    The left column varies eccentricity and inclination angle $\iota$, 
    while the right column varies eccentricity and aligned effective spin $\chi_{\rm eff}$. 
    The colour scale indicates the sensitive volume in $\mathrm{Gpc}^3$. 
    A reduced sensitive volume is visible for the $12$--$12\,M_\odot$ inclination case across the 
    edge-on region, $\iota \in [1.25, 1.75]$, relative to the face-on and face-off orientations; 
    this is primarily an inclination effect, reflecting the weaker observed strain amplitude of 
    edge-on binaries, and is present across the eccentricity range rather than driven by eccentricity.
    For the $30$--$30\,M_\odot$ case the edge-on deficit is weaker overall but not absent, 
    appearing mainly in the high-eccentricity corner (cf.\ Figure~\ref{fig:inc_ecc_diagno}). 
    The aligned-spin panels show no clear systematic reduction in sensitive volume 
    with increasing eccentricity or spin.
    }
    \label{fig:sensitivity_volume_params}
\end{figure*}

\begin{figure*}[t]
    \centering
    \includegraphics[width=\textwidth]{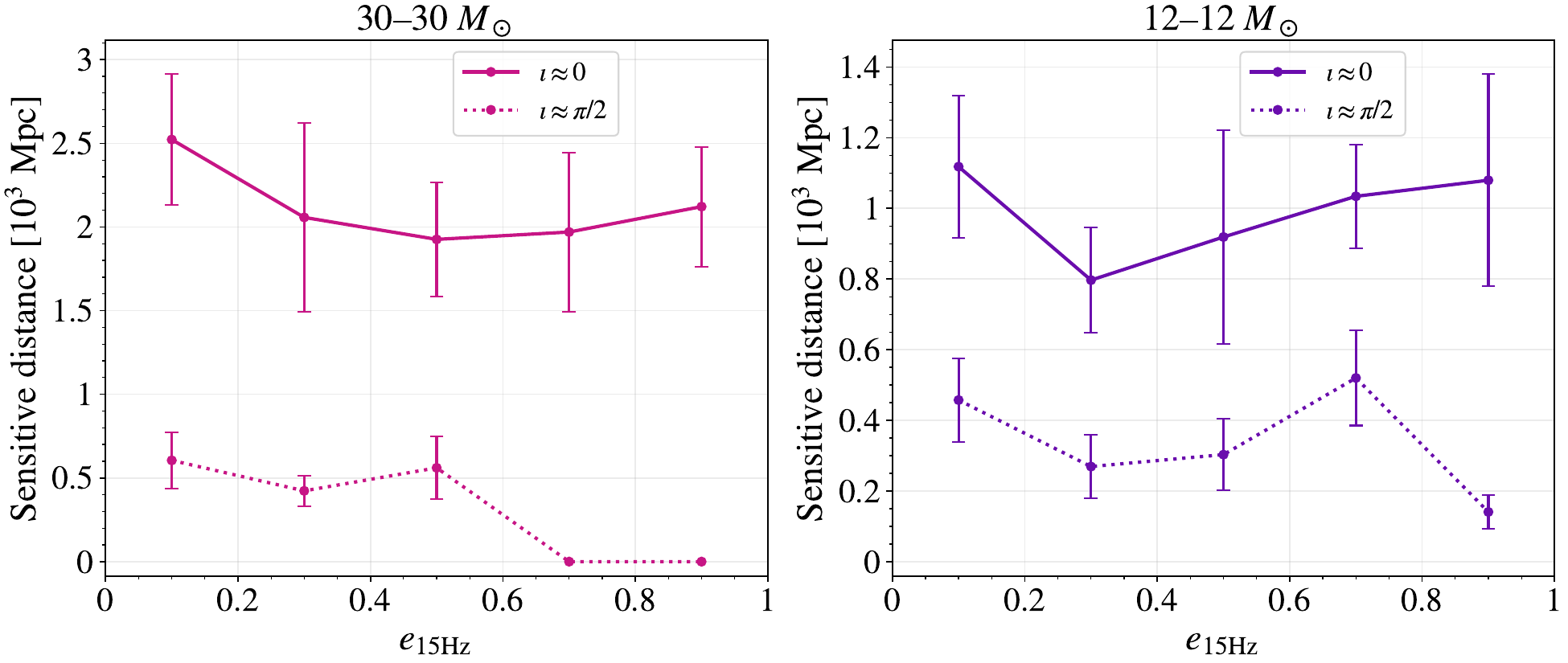}
    \caption{
    Average sensitive distance as a function of orbital eccentricity $e_{15\rm Hz}$ for 
    two representative inclinations from injection campaign~5, 
    evaluated at $\mathrm{FAR}_{\rm threshold} = 12.1\,{\rm yr}^{-1}$.
    The two panels show the $30$--$30\,M_\odot$ (left) and $12$--$12\,M_\odot$ (right) configurations; 
    within each panel, the solid and dotted curves show the face-on ($\iota \approx 0$) and 
    edge-on ($\iota \approx \pi/2$) orientations respectively.
    Error bars quantify the uncertainty in the estimated sensitive distance, 
    computed as in Figure~\ref{fig:avg_snr}.
    For the near-edge-on ($\iota \approx \pi/2$) case the higher-mass population 
    shows a marked decline of sensitive distance with eccentricity, 
    falling essentially to zero at the highest eccentricities, 
    whereas the lower-mass population remains comparatively flat.
    }
    \label{fig:inc_ecc_diagno}
\end{figure*}

\subsection{Injection campaign 5: sensitivity volume with varying eccentricity, chirp mass, inclination, and spin} \label{subsec:results_campaign5}

Injection campaign~5 extends the previous fixed-configuration studies by 
examining the sensitive volume over a broader source-parameter space. 
Recall that this campaign aims to answer whether eccentricity affects the sensitivity volume 
when it is varied together with chirp mass, inclination, and aligned effective spin. 
The sensitive volume and distance reported below are estimated as in 
Appendix~\ref{appendix:sensitive_volume}.

Figure~\ref{fig:sensitivity_volume_cmass} shows the sensitive volume 
as a function of eccentricity and chirp mass, evaluated at a false-alarm-rate 
threshold of $\mathrm{FAR}_{\rm threshold} = 12.1\,{\rm yr}^{-1}$, i.e.\ one false alarm per $30$ 
days --- the standard \gstlal{} detection threshold, 
and the same found-injection criterion used for Campaigns~1--4 (Table~\ref{tab:injection_scheme}). 
The behaviour separates cleanly by chirp mass. 
For $\mathcal{M}_c \gtrsim 30\,M_\odot$, the sensitive volume shows essentially no dependence on eccentricity. 
Below $\mathcal{M}_c \approx 30\,M_\odot$, however, a slight but systematic trend of decreasing sensitive 
volume with increasing eccentricity emerges. 
This mass-dependent trend aligns with the sensitive-distance results 
(Figure~\ref{fig:sensitive_distance_far12}), 
where the same lower-chirp-mass bins exhibit a clear reduction with eccentricity while 
the high-mass bin remains flat. 
It is also consistent with Ref.~\cite{Ramos-Buades:2020eju}, 
which found that eccentricity affects the search sensitivity, and 
hence the search sensitive distance, in the lower-mass regime.

To quantify this mass-dependent behaviour more directly, 
Figure~\ref{fig:sensitive_distance_far12} shows the sensitive distance as a function of eccentricity 
for three chirp-mass bins, again at $\mathrm{FAR}_{\rm threshold} = 12.1\,{\rm yr}^{-1}$; 
the error bars give the uncertainty on each estimate, 
computed as described in the caption. 
For the high-mass bin ($30.44 \le \mathcal{M}_c \le 40.44\,M_\odot$), 
the sensitive distance is essentially flat with eccentricity, 
decreasing by less than $20\%$ even at the highest eccentricities, 
so the effect of eccentricity is negligible in this range. 
For the two lower-mass bins 
($10.44 \le \mathcal{M}_c < 20.44\,M_\odot$ and $20.44 \le \mathcal{M}_c < 30.44\,M_\odot$), 
the sensitive distance decreases systematically with eccentricity, 
reaching a reduction of roughly $50\%$ and $40\%$, respectively, 
by $e_{15\rm Hz} \approx 0.75$ relative to the quasicircular case, 
a decline that exceeds the error bars on the individual points. 
To put this number into context, 
the sensitivity increase in binary distance due to the  
the massive effort in detector upgrades installed before fourth Observing run (O4) was around $20-50\%$~\cite{PhysRevD.111.062002}. 

The corresponding sensitive-volume results, 
evaluated at $\mathrm{FAR}_{\rm threshold} = 12.1\,{\rm yr}^{-1}$, 
are presented in Figure~\ref{fig:sensitivity_volume_params} when eccentricity is 
varied together with inclination or aligned effective spin for 
the $12$--$12\,M_\odot$ and $30$--$30\,M_\odot$ mass configurations. 
In the inclination panels, the $12$--$12\,M_\odot$ configuration shows 
a reduced sensitive volume across the edge-on band $\iota \in [1.25, 1.75]$ (around $\iota \simeq \pi/2$) 
relative to the face-on and face-off orientations; 
this is an inclination effect, 
present across the eccentricity range rather than driven by eccentricity itself. 
This behaviour is qualitatively similar to Figure~12 of Ref.~\cite{Gadre:2024ndy}, 
and is expected because edge-on binaries have a weaker observed strain amplitude than 
face-on or face-off binaries. 
For the $30$--$30\,M_\odot$ configuration the edge-on deficit is weaker overall, 
but it is not absent: the $30$--$30\,M_\odot$ edge-on sensitive distance falls 
specifically in the high-eccentricity corner (Figure~\ref{fig:inc_ecc_diagno}), 
whereas the $12$--$12\,M_\odot$ edge-on distance is already low and stays comparatively flat with 
eccentricity. 
The two figures therefore describe different axes, where the heatmap the variation across inclination, 
the line plot the variation with eccentricity within the edge-on slice, and 
are consistent once read that way.

Figure~\ref{fig:inc_ecc_diagno} makes this inclination dependence explicit by 
showing the estimated sensitive distance as a function of eccentricity for two representative inclinations, 
face-on ($\iota \approx 0$) and edge-on ($\iota \approx \pi/2$). 
For the face-on orientation, the loudest configuration, 
both mass populations remain essentially flat with eccentricity, scattering within their error bars with 
no significant trend, as expected for signals that stay well above the detection threshold across 
the eccentricity range. 
For nearly edge-on injections ($\iota \approx \pi/2$), by contrast, the $30$--$30\,M_\odot$ population 
exhibits a clear decline in estimated sensitive distance with increasing eccentricity, 
whereas the $12$--$12\,M_\odot$ population remains comparatively flat. 
In the higher-mass sample, this decline is accompanied by a falling fraction of 
matched injections passing the \gls{FAR} threshold of $12.1\,\mathrm{yr}^{-1}$ as eccentricity increases, 
dropping to essentially zero in the highest-eccentricity bin. 
This indicates a reduced search acceptance at high eccentricity for near-edge-on, 
higher-mass systems. 
However, the available diagnostics do not uniquely attribute this reduction to waveform mismatch, 
as edge-on binaries already have a weak observed strain and 
the small number of passing injections increases the statistical uncertainty.

In the aligned-spin panels, the sensitive volume remains broadly comparable across 
eccentricity and spin values for both mass configurations. 
Although local fluctuations are present, these do not form a clear systematic trend 
indicating that high eccentricity or aligned spin strongly reduces the sensitive volume. 
Overall, the results of injection campaign~5 are broadly consistent with those of 
the previous four campaigns: eccentricity does not generally lead to a strong reduction in 
sensitive volume within the parameter space considered here. 
The main exception is the $12$--$12\,M_\odot$ inclination-dependent case near edge-on orientation, 
where the sensitive volume decreases noticeably. 
This suggests that matched-filtering searches remain capable of recovering eccentric signals with 
broadly comparable sensitivity to quasicircular injections, 
while lower-mass, near-edge-on systems represent a more challenging region of parameter space.

\section{Conclusion} \label{sec:conclusion}

In this work we have investigated how orbital eccentricity affects the detectability of 
binary black holes in matched-filtering gravitational-wave searches. 
Using the \gstlal{} pipeline with \dali{} eccentric injections recovered by 
a standard quasicircular template bank, we carried out five injection campaigns that span eccentricity, 
chirp mass, mass ratio, inclination and aligned spin. 
In each campaign we measured the recovered \gls{SNR}, 
the signal-consistency statistic, the false-alarm rate and the sensitive volume and distance. 
The aim was to determine whether eccentric signals are lost by current quasicircular searches and 
hence whether dedicated eccentric template banks are required.

Our central finding is that high-mass eccentric \glspl{BBH} remain essentially 
as detectable as 
their quasicircular counterparts up to the highest eccentricities probed, 
except near edge-on orientation, 
while a measurable loss in search sensitivity emerges primarily for lower-mass systems. 
For high-mass binaries ($\mathcal{M}_c \gtrsim 30\,M_\odot$), the recovered \gls{SNR} and 
found fraction remain independent of eccentricity up to $e_{\rm 15Hz}=0.99$, and 
the sensitive volume and sensitive distance up to $e_{\rm 15Hz}=0.9$, 
apart from a decline confined to the high-eccentricity, 
near-edge-on corner that rests on a small number of recovered injections. 

In contrast, for lower-mass systems ($\mathcal{M}_c \lesssim 30\,M_\odot$), 
the sensitive distance decreases with eccentricity by up to approximately 
$50\%$ by $e_{\rm 15Hz}\approx0.75$. 
This reduction in terms of volume would render nearly 90\% of the signals, that 
could be detected if eccentricity were included, undetectable, and is comparable 
in the loss of sensitivity to some of the improvements in binary 
neutron star sensitivity distance achieved between observing runs in the past. 

This work extends previous studies by probing substantially higher eccentricities 
and a broader source-parameter space using end-to-end recovery, 
using the \gstlal{} pipeline to measure the actual search sensitivity. 
Covering orbital eccentricities up to $e_{\rm 15Hz}=0.99$, 
our results demonstrate for the first time in this regime that a large waveform mismatch 
does not necessarily translate into a comparable pipeline-level loss in search sensitivity. 
In particular, the absence of a measurable eccentricity-dependent sensitivity loss for 
high-mass binaries indicates that dedicated eccentric templates, or 
an additional eccentricity dimension in the template bank, 
may not be necessary for preserving search sensitivity in this region of parameter space. 
At lower masses, dedicated eccentric templates could improve recovery, but 
their potential gain must be weighed against the computational cost of expanding the 
template bank and the comparatively modest absolute increase in detectable reach.

This discrepancy between the preliminary waveform-level prediction and the full injection results has two origins. First, the preliminary study compares each eccentric signal with the quasicircular waveform of \emph{identical} source parameters, whereas the search maximises over the whole template bank and can recover the signal with a better-matching, shifted-parameter template (Figure~\ref{fig:wf_overlap}); part of the apparent loss is therefore simply the gap between the same-parameter match and the bank fitting factor, a well-known quantity rather than a pipeline effect. Second, and more importantly, the end-to-end matched-filtering search is considerably more robust than either overlap-based proxy would suggest, because matched-filtering pipelines do not rank triggers using the recovered \gls{SNR} alone; instead, the ranking statistic also incorporates the signal-consistency statistic $\xi^2$, allowing the search to favour triggers that provide a better balance between signal strength and waveform consistency. Moreover, the signal-consistency statistic $\xi^2$ is evaluated only within a short time window around merger. For lower-mass eccentric binaries, much of the eccentricity-induced waveform structure is concentrated earlier in the inspiral, while the orbit becomes progressively more circular toward merger. Consequently, the SNR time series within the signal-consistency $\xi^2$ window can remain sufficiently similar to that of a quasicircular template, limiting the signal-consistency penalty. This helps explain why the loss in pipeline-level sensitivity is substantially smaller than would be inferred from waveform match alone.

Several extensions follow naturally from this work. 
It is important to extend the analysis to lower total masses, 
progressing from \glspl{BBH} towards neutron-star--black-hole and binary-neutron-star systems. 
In these regimes, the longer in-band inspirals allow eccentricity-induced modulations to 
accumulate over many more cycles, potentially making eccentricity more important for matched-filter recovery. 
Such systems also introduce additional physical effects, 
including tidal interactions and different eccentric-source populations, 
which may further influence detectability. 
Another important direction is to quantify directly the cost and benefit of constructing a 
dedicated eccentric template bank in the low-mass and near-edge-on regions identified in this work. 
This would allow the potential gain in search sensitivity to be assessed against the substantial 
computational cost associated with expanding the template-bank parameter space. 
Further extensions could also consider a broader range of source configurations, 
such as unequal mass ratios and more general spin configurations, 
to determine whether the conclusions found here remain valid beyond the equal-mass, 
aligned-spin or non-spinning systems considered in the present campaigns.

\section{Acknowledgement}

The authors would like to thank Alan J. Weinstein, Chun Yu Kenny Ng and Ming Chung Chu for their thoughtful and insightful comments on this work. The authors are also grateful to Kipp Cannon and Heather Fong for many helpful discussions and suggestions throughout the project.
This work was supported by the Research Grants Council of Hong Kong (Project Nos.\ CUHK 14304622, 14307923 and 14307724) and by the Direct Grant for Research from the Research Committee of The Chinese University of Hong Kong.
J.C.L.C. acknowledges support from the Villum Investigator program supported by the VILLUM Foundation (grant no. VIL37766 and no. VIL53101) and the DNRF Chair program (grant no. DNRF162) by the Danish National Research Foundation. A.~Y.~A.~L. would like to acknowledge the support from the Croucher Foundation.
The authors acknowledge the use of OpenAI's ChatGPT and Anthropic's Claude for proofreading, language refinement, and suggestions on the wording and organization of the manuscript.

This work made use of data obtained from the Gravitational Wave Open Science Center (\url{https://gwosc.org}), a resource provided by the LIGO Laboratory, the LIGO Scientific Collaboration, the Virgo Collaboration and KAGRA. This material is based upon work supported by NSF's LIGO Laboratory which is a major facility fully funded by the National Science Foundation. The Virgo detector is supported by the French CNRS, the Italian INFN and the Dutch Nikhef. KAGRA is supported by MEXT and JSPS in Japan, together with the NRF in Korea and Academia Sinica in Taiwan.

The authors are grateful for computational resources provided by the LIGO Laboratory and supported by National Science Foundation Grants No.\ PHY-0757058 and No.\ PHY-0823459.

\newpage

\appendix

\section{Overview of \dali}\label{appendix:dali}

\dali{}~\cite{Chiaramello:2020ehz,Nagar:2024dzj} is the eccentric extension of the \texttt{TEOBResumS} family of \gls{EOB} waveform models~\cite{Nagar:2018zoe}. The \gls{EOB} formalism maps the general-relativistic two-body problem onto the motion of a single effective particle in a deformed background metric, with the deformation controlled by the symmetric mass ratio $\nu = m_1 m_2 / M^2$, where $M = m_1 + m_2$~\cite{Buonanno:1998gg,Buonanno:2000ef}. By resumming the \gls{PN} expansion within this effective description and calibrating the remaining free functions to \gls{NR} simulations, the model produces complete \gls{IMR} waveforms. \dali{} generalizes this construction from quasicircular to generic planar orbits, spanning eccentric and dynamical-capture (hyperbolic) configurations within a single framework. In this work we use \dali{} to generate the eccentric injections described in Section~\ref{subsec:generate_wf}.

Two parameters characterize the eccentric orbit in addition to the masses and aligned spins. The \emph{eccentricity} $e$ quantifies the departure of the orbit from circularity: $e = 0$ corresponds to a quasicircular orbit, while $0 < e < 1$ describes a bound eccentric orbit. Because eccentricity is not uniquely defined in general relativity, \dali{} fixes it through the \gls{EOB} initial data at a chosen reference frequency. Writing the orbital separation in terms of the semi-latus rectum $p$ and the relativistic (true) anomaly $\zeta$,
\begin{align}
    r = \frac{p}{1 + e \cos\zeta},
\label{eq:dali_orbit}
\end{align}
the periastron and apastron separations are $r_{-} = p/(1+e)$ and $r_{+} = p/(1-e)$, respectively. The \emph{mean anomaly} $\ell$ is the phase-like variable that specifies the location of the bodies along the eccentric orbit at the reference time. It increases uniformly between successive periastron passages and vanishes at periastron. Following the convention adopted in Section~\ref{subsec:generate_wf}, our injections are initialized at $f_{\rm ecc} = 15\,{\rm Hz}$ with the orbit set to periastron.

\subsection{EOB dynamics and resummation}

The conservative dynamics are governed by the real \gls{EOB} Hamiltonian,
\begin{align}
    H_{\rm EOB} = M\sqrt{1 + 2\nu\left(\hat{H}_{\rm eff} - 1\right)},
\label{eq:dali_HEOB}
\end{align}
where $\hat{H}_{\rm eff}$ is the effective Hamiltonian in units of the reduced mass $\mu = \nu M$. For nonspinning binaries its orbital part reads
\begin{align}
    \hat{H}_{\rm eff} = \sqrt{p_{r_*}^{2} + A(u)\left(1 + p_\varphi^{2}\,u^{2} + z_3\, p_{r_*}^{4}\right)},
\label{eq:dali_Heff}
\end{align}
where $u = M/r$, $p_\varphi$ and $p_{r_*}$ are the (dimensionless) azimuthal and tortoise radial momenta, and $z_3 = 2\nu(4 - 3\nu)$. The radial potential $A(u)$ encodes the conservative interaction. Its \gls{PN}-expanded form is not used directly; instead it is \emph{resummed} with a Pad\'e approximant, which greatly improves its behaviour in the strong-field regime near merger~\cite{Damour:2008gu,Nagar:2018zoe},
\begin{align}
    A(u) = P^{n}_{m}\!\left[A_{\rm PN}(u)\right].
\label{eq:dali_Apade}
\end{align}

The dynamics are driven by the gravitational \gls{RR} through Hamilton's equations, which for the momenta take the form
\begin{align}
    \dot{p}_\varphi &= \hat{\mathcal{F}}_\varphi, \label{eq:dali_pphidot}\\
    \dot{p}_{r_*} &= -\sqrt{\frac{A}{B}}\,\partial_r \hat{H}_{\rm EOB} + \hat{\mathcal{F}}_{r_*}, \label{eq:dali_prdot}
\end{align}
where $\hat{\mathcal{F}}_\varphi$ and $\hat{\mathcal{F}}_{r_*}$ are the azimuthal and radial components of the \gls{RR} force and $B(u)$ is the second radial potential. For quasicircular orbits the radial force $\hat{\mathcal{F}}_{r_*}$ vanishes and only $\hat{\mathcal{F}}_\varphi$ is retained. For eccentric orbits, however, $\hat{\mathcal{F}}_{r_*}$ can no longer be neglected, and its inclusion is one of the essential ingredients of \dali{}~\cite{Chiaramello:2020ehz,Nagar:2024dzj}. In the model, the leading (Newtonian) prefactor entering the waveform multipoles and the azimuthal force is generalized from its circular form to a general-orbit expression obtained by explicitly taking the time derivatives of the Newtonian mass and current multipole moments, thereby retaining the full radial dependence of the eccentric motion; the higher-\gls{PN} eccentric corrections to the radial force are additionally Pad\'e-resummed.

\subsection{Factorized and resummed waveform}

The waveform multipoles $h_{\ell m}$ are constructed using the factorized, resummed form~\cite{Damour:2008gu},
\begin{align}
    h_{\ell m} = h_{\ell m}^{(N,\epsilon)}\,\hat{S}_{\rm eff}^{(\epsilon)}\,T_{\ell m}\,e^{i\delta_{\ell m}}\,\left(\rho_{\ell m}\right)^{\ell},
\label{eq:dali_hlm}
\end{align}
where $h_{\ell m}^{(N,\epsilon)}$ is the Newtonian prefactor, $\hat{S}_{\rm eff}^{(\epsilon)}$ is the effective source (the \gls{EOB} energy or angular momentum depending on the parity $\epsilon$ of the mode), $T_{\ell m}$ is the resummed tail factor accounting for the back-scattering of the radiation off the background curvature, $\delta_{\ell m}$ is a residual phase, and $\rho_{\ell m}$ is the residual amplitude. Raising $\rho_{\ell m}$ to the power $\ell$ and further resumming it (typically with a Pad\'e approximant) is what makes the factorized waveform accurate deep into the strong-field regime. For eccentric orbits, the Newtonian prefactor $h_{\ell m}^{(N,\epsilon)}$ is replaced by the general-orbit expression discussed above, so that the waveform naturally reproduces the spectrum of orbital harmonics excited by the radial motion. The azimuthal \gls{RR} force is in turn obtained from the resummed multipolar fluxes,
\begin{align}
    \hat{\mathcal{F}}_\varphi = -\frac{1}{\nu\,\Omega}\,\frac{1}{16\pi}\sum_{\ell,m}\left(m\,\Omega\right)^{2}\left|h_{\ell m}\right|^{2},
\label{eq:dali_flux}
\end{align}
where $\Omega = \dot{\varphi}$ is the orbital frequency, so that the same resummed multipoles enter both the emitted waveform and the dissipative dynamics.

In summary, the accuracy of \dali{} rests on three resummations that carry the model from the quasicircular to the eccentric regime: (i) the radial potential $A(u)$ from Eq.~\eqref{eq:dali_Apade}, (ii) the residual multipole amplitudes $\rho_{\ell m}$ and the tail factor $T_{\ell m}$ in the factorized waveform from Eq.~\eqref{eq:dali_hlm}, and (iii) the higher-\gls{PN} eccentric part of the radial radiation-reaction force. The merger--ringdown is attached through an \gls{NR}-informed description of the quasi-normal-mode ringing. When compared against publicly available eccentric \gls{NR} simulations, \dali{} reaches an unfaithfulness at the percent level~\cite{Chiaramello:2020ehz,Bonino:2024xrv}. The dependence of the recovered match and optimal \gls{SNR} on the eccentricity $e_{\rm 15Hz}$ and the mean anomaly, for signals generated with a starting frequency of $15\,{\rm Hz}$, is examined in the preliminary study of Section~\ref{subsec:prelim_match}.

\section{The \gstlal{} search pipeline}\label{appendix:gstlal}
We provide a complementary description of the \gstlal{} matched-filtering search pipeline we employed in this paper. 
We introduce the autocorrelation-based signal-consistency test, and the likelihood-ratio ranking statistic. 

\subsection{Autocorrelation based signal-consistency test} \label{appendix:signal_consistency_test}

When the matched-filter \gls{SNR} time series, $\rho(t)$, exceeds a predefined threshold of $\rho=4$, the search pipeline records the corresponding template parameters as a candidate event, known as a trigger~\cite{Messick:2016aqy}.
Non-Gaussian noise transients (``glitches’’) can generate high-\gls{SNR} triggers that mimic \gls{GW} signals. 
To assess the consistency of a trigger with the expected signal morphology, \gstlal{} computes a signal-consistency statistic, $\xi^{2}$, derived from the template autocorrelation function. This statistic measures the deviation between the observed matched-filter \gls{SNR} time series and that predicted by the autocorrelation of the associated template.

For a given template $h_{T,j}$, with the index $j$ labelling the template and the two components $h_{T,2j}$ and $h_{T,2j+1}$ denoting its two orthogonal phase quadratures, the signal-consistency statistic is written in terms of the complex matched-filter \gls{SNR} time series $z_j(t)$ of that template as
\begin{align}
	\xi_{j}^{2} = \frac{\int_{-\delta t}^{\delta t} \left|z_{j}(t)-z_{j}(0)R_{j}(t)\right|^{2} dt}{\int_{-\delta t}^{\delta t} 2 - 2\left|R_{j}(t)\right|^{2}dt},
\label{eq:search2}
\end{align}
where
\begin{align}
	R_j(t) = \int \frac{|\tilde{h}_{T, \; 2j}(f)|^2 + |\tilde{h}_{T, \; 2j+1}(f)|^2}{S_n(f)} e^{2\pi i f t} df,
\label{eq:search2-1}
\end{align}
denotes the expected \gls{SNR} time series predicted by the template autocorrelation, with $S_n(f)$ the noise \gls{PSD} defined in Section~\ref{subsec:wf_overlapping}.
The reference time  $t=0$ is defined as the location of the global maximum of the observed matched-filter \gls{SNR} time series. The tunable parameter $\delta t$ specifies the width of the time window centered on the peak within which the integration is performed~\cite{Messick:2016aqy}.
The statistic $\xi_j^2$ is normalized such that a perfect match between the observed and expected \gls{SNR} time series yields $\xi_j^2 = 1$.
Any mismatch between the two time series increases the value of $\xi_j^2$, indicating a departure from the signal morphology predicted by the template.

As an illustrative extreme case, consider a glitch whose matched-filter output is a scaled delta function, $z_j(0)\delta(t)$, at $t=0$. 
The resulting \gls{SNR} time series is then zero throughout the integration window except at the peak time, whereas the expected \gls{SNR} time series, $z_j(0)R_j(t)$, retains the characteristic autocorrelation structure of the template. 
This large mismatch causes the numerator of $\xi_j^2$ to increase substantially. Since the discrepancy scales with the amplitude $z_j(0)$, louder noise transients generally yield larger values of $\xi_j^2$. 
Similarly, astrophysical signals that deviate from the template waveform may also alter the signal-consistency statistic $\xi^2$, although the impact of such waveform mismatches is generally more subtle and nontrivial~\cite{Bondarescu:2022srx}.

\subsection{Likelihood-ratio ranking statistic}\label{subsec:gstlal_likelihood}
For each trigger, \gstlal{} computes a likelihood-ratio ranking statistic using a set of detection statistics, such as the matched-filter SNR $\rho$ and the signal-consistency statistic $\xi^2$~\cite{Messick:2016aqy,Sachdev:2019vvd,Cannon:2015gha}. 
This statistic is defined as the ratio of the probability of observing the trigger’s detection statistics under the signal hypothesis to that under the noise hypothesis, thereby quantifying the degree to which the trigger is consistent with a \gls{GW} signal rather than instrumental noise.
 
The likelihood-ratio ranking statistic is given by

\begin{align}
	\mathcal{L} = \frac{P(\vec{D}_H,\vec{O},\vec\rho,\vec\xi^2,\left[\vec{\Delta t}, \vec{\Delta \phi}\right]|\vec\theta,\mathrm{signal})}{P(\vec{D}_H,\vec{O},\vec\rho,\vec\xi^2,\left[\vec{\Delta t}, \vec{\Delta \phi}\right]|\vec\theta,\mathrm{noise})}\cdot\frac{P(\vec\theta|\mathrm{signal})}{P(\vec\theta|\mathrm{noise})},
\label{eq:likelihood_ratio_ranking}
\end{align}

where $\vec{D}_H$ denotes the detector horizon distances, $\vec{O}$ identifies the set of detectors that contributed to the trigger, and $\vec{\theta}$ represents the template parameters. 
The vectors $\vec{\rho}$ and $\vec{\xi}^{2}$ contain the matched-filter SNRs and autocorrelation-based signal-consistency statistics, respectively, measured in the participating detectors. 
The quantities [$\vec{\Delta t}$, $\vec{\Delta \phi}$] denote the relative arrival times and coalescence phases between detector pairs, which provide information on the coherence of the candidate signal across the detector network. 
\gstlal{} subsequently constructs the complementary cumulative distribution function of the log-likelihood ratio,
\begin{align}
    \mathcal{C}(\ln \mathcal{L}^* | \text{noise}) = \int_{\ln \mathcal{L}^*}^\infty P\left(\ln \mathcal{L} | \text{noise}\right) d\ln\mathcal{L},
\end{align}
which gives the probability that a noise trigger attains a log-likelihood ratio greater than or equal to a threshold $\ln \mathcal{L}^*$.
Using this distribution, the False-Alarm Probability (FAP), or the $p$-value can be computed. 
Specifically, consider a set of $M$ independent coincident noise-like events ($\{N_1, N_2, ... , N_M\}$.
The probability that at least one of these events has a log-likelihood ratio exceeding $\ln\mathcal{L}^*$ is:

\begin{align}
    P(\ln\mathcal{L}\geq \ln\mathcal{L}^* | N_1, ... ,N_M) &= 1 - \left(e^{-\mathcal{C}(\ln \mathcal{L}^* | \text{noise})}\right)^M\\
    &= 1 - e^{-M\mathcal{C}(\ln \mathcal{L}^* | \text{noise})}.
\end{align}

Here, $e^{-\mathcal{C}(\ln \mathcal{L}^* | \text{noise})}$ corresponds to the probability that a Poisson process with mean $\mathcal{C}(\ln \mathcal{L}^* | \text{noise})$ produces no events with $\ln\mathcal{L} \geq \ln \mathcal{L}^*$.
Consequently, the FAP quantifies the probability that noise alone generates at least one trigger with a likelihood ratio greater than or equal to that of the candidate under consideration, $\mathcal{L}^*$. 
Since astrophysical \gls{GW} signals are expected to yield larger likelihood ratios than noise fluctuations, smaller FAP values indicate a higher probability that a candidate is of astrophysical origin.
A closely related quantity is the \gls{FAR}, which is defined as
\begin{align}\label{eq:FAR}
    \text{FAR} = \frac{\mathcal{C}(\ln\mathcal{L}^*|\text{noise})}{T}
\end{align}
where $T$ denotes the total observing time used to estimate the background distribution. 
The \gls{FAR} represents the expected rate at which noise produces triggers with likelihood ratios greater than or equal to $\mathcal{L}^*$. 
As with the FAP, lower \gls{FAR} values correspond to a higher confidence that a candidate is a genuine \gls{GW} signal~\cite{Cannon:2015gha,Messick:2016aqy,Sachdev:2019vvd}.

For \gstlal{}, a commonly adopted threshold for confidently identifying a \gls{GW} event is a \gls{FAR} of $3.85\times10^{-7}\,\mathrm{Hz}$, corresponding to approximately one expected false alarm every 30 days.

\section{Estimation of the sensitive volume and distance}\label{appendix:sensitive_volume}

The sensitivity of the search to a given source configuration is quantified by the sensitive volume, the spatial volume within which sources are recovered by the pipeline. In the Euclidean (non-cosmological) approximation appropriate to the distances probed here, the sensitive volume in a given source-parameter bin is
\begin{equation}
    V = \int_0^\infty \varepsilon(r)\,4\pi r^2\,dr ,
    \label{eq:sensvol_def}
\end{equation}
where $r$ is the luminosity distance and $\varepsilon(r)$ is the detection efficiency, i.e.\ the fraction of injections at distance $r$ that are recovered by the search.

We evaluate Eq.~\eqref{eq:sensvol_def} by Monte-Carlo integration over the injection set. The injections in each bin are distributed uniformly in luminosity distance over $[r_{\min}, r_{\max}]$, so that each injection carries equal weight and the estimator reduces to
\begin{equation}
    V = \frac{4\pi\,\Delta r}{N_{\rm inj}} \sum_{i \in \mathrm{found}} r_i^2 ,
    \qquad \Delta r = r_{\max} - r_{\min} ,
    \label{eq:sensvol_mc}
\end{equation}
where $N_{\rm inj}$ is the total number of injections in the bin, $r_i$ is the luminosity distance of the $i$-th injection, and the sum runs over the injections that are recovered (``found''). An injection is counted as found when the \gstlal{} false-alarm rate of its recovered trigger falls below the detection threshold $\mathrm{FAR}_{\rm threshold} = 12.1\,{\rm yr}^{-1}$ (equivalently $3.85\times10^{-7}\,{\rm Hz}$; see Appendix~\ref{subsec:gstlal_likelihood}), the same criterion used for the injection-recovery counts of Table~\ref{tab:injection_scheme}. We quote the equivalent sensitive distance, defined as the radius of the sphere of the same volume,
\begin{equation}
    D = \left( \frac{3V}{4\pi} \right)^{1/3} .
    \label{eq:sensdist}
\end{equation}
The error bars shown on the sensitive volume and distance reflect the Monte-Carlo counting uncertainty arising from the finite number of found injections in each bin.

The estimator is evaluated with the same noise \gls{PSD} and injection-distance distribution for the eccentric and quasicircular configurations. Systematic effects from the Euclidean approximation and from the antenna response therefore largely cancel in the relative comparison between them, so that the \emph{eccentricity dependence} of $V$ and $D$ is robust even where their absolute normalization is only approximate.

\clearpage
\bibliography{citations}

\end{document}